\documentclass[english,preprint,tightenlines,eqsecnum,nofootinbib]{revtex4}
\usepackage[T1]{fontenc}
\usepackage[latin9]{inputenc}
\usepackage{babel}
\usepackage{amsmath}
\usepackage{amssymb}
\usepackage{graphicx}
\usepackage[unicode=true,pdfusetitle,
 bookmarks=true,bookmarksnumbered=false,bookmarksopen=false,
 breaklinks=false,pdfborder={0 0 1},backref=false,colorlinks=false]
 {hyperref}

\makeatletter
\@ifundefined{textcolor}{}
{%
 \definecolor{BLACK}{gray}{0}
 \definecolor{WHITE}{gray}{1}
 \definecolor{RED}{rgb}{1,0,0}
 \definecolor{GREEN}{rgb}{0,1,0}
 \definecolor{BLUE}{rgb}{0,0,1}
 \definecolor{CYAN}{cmyk}{1,0,0,0}
 \definecolor{MAGENTA}{cmyk}{0,1,0,0}
 \definecolor{YELLOW}{cmyk}{0,0,1,0}
}

\usepackage{braket}
\usepackage{caption}
\@ifundefined{showcaptionsetup}{}{%
 \PassOptionsToPackage{caption=false}{subfig}}
\usepackage{subfig}
\makeatother

\begin{document}

\title{Mode-sum regularization of $\left\langle \phi^{2}\right\rangle $
in the angular-splitting method}

\author{Adam Levi and Amos Ori}

\address{Department of physics, Technion-Israel Institute of Technology,\\
Haifa 3200, Israel }
\begin{abstract}
The computation of the renormalized stress-energy tensor or $\left\langle \phi^{2}\right\rangle _{ren}$
in curved spacetime is a challenging task, at both the conceptual
and technical levels. Recently we developed a new approach to compute
such renormalized quantities in asymptotically-flat curved spacetimes,
based on the point-splitting procedure. Our approach requires the
spacetime to admit some symmetry. We already implemented this approach
to compute $\left\langle \phi^{2}\right\rangle _{ren}$ in a stationary
spacetime using $t$\textit{-splitting}, namely splitting in the time-translation
direction. Here we present the \textit{angular-splitting }version
of this approach, aimed for computing renormalized quantities in a
general (possibly dynamical) spherically-symmetric spacetime. To illustrate
how the angular-splitting method works, we use it here to compute
$\left\langle \phi^{2}\right\rangle _{ren}$ for a quantum massless
scalar field in Schwarzschild background, in various quantum states
(Boulware, Unruh, and Hartle-Hawking states). We find excellent agreement
with the results obtained from the \emph{$t$}\textit{\emph{-splitting}}
variant, and also with other methods. Our main goal in pursuing this
new mode-sum approach was to enable the computation of the renormalized
stress-energy tensor in a dynamical spherically symmetric background,
e.g. an evaporating black hole. The angular-splitting variant presented
here is most suitable to this purpose.
\end{abstract}
\maketitle

\section{Introduction}

The dynamical process of black-hole (BH) evaporation attracts much
interest since Hawking's discovery that BHs emit radiation \cite{Hawking - Particle creation by black holes}.
This is because the BH evaporation phenomenon is intimately related
to the connection between gravity and quantum mechanics. The main
theoretical framework that allows us to study this process is semiclassical
gravity. In this framework one considers a classical curved metric
$g_{\alpha\beta}\left(x\right)$ with a quantum field. For simplicity
we shall consider here a scalar field $\phi\left(x\right)$. This
quantum field evolves according to Klein-Gordon equation
\begin{equation}
\left(\square-m^{2}-\xi R\right)\phi=0,\label{eq: Intro - KG eq.}
\end{equation}
where $m$ is the field's mass \footnote{Throughout most of this paper, however, we shall consider the massless
case $m=0.)$} and $\xi$ is its curvature coupling. The metric evolves according
to the semiclassical Einstein equation
\begin{equation}
R_{\alpha\beta}-\frac{1}{2}Rg_{\alpha\beta}=8\pi\left\langle T_{\alpha\beta}\right\rangle _{ren},\label{eq: Inreo - SC Einstein eq.}
\end{equation}
where $R_{\alpha\beta}$ and $R$ are the Ricci tensor and Ricci scalar,
and $\left\langle T_{\alpha\beta}\right\rangle _{ren}$ is the renormalized
expectation value of the stress-energy tensor of the quantum field.
It is constructed from the fields' modes, and it depends on the metric
$g_{\alpha\beta}\left(x\right)$ (and on the field's quantum state).
Throughout this paper we use relativistic units $c=G=1$.

All the above is known for  four decades, and yet so far no one was
able to solve these two coupled equations and provide detailed quantitative
description of the semiclassical evaporation process. The main reason
is that computation of the renormalized stress tensor turns out to
be extremely difficult in a general curved background \textemdash{}
even if the background metric $g_{\alpha\beta}\left(x\right)$ is
prescribed. This difficulty emerges from the regularization process.
Much like in flat spacetime, the ``bare'' expectation value of the
stress tensor is divergent; nevertheless, in flat spacetime this divergence
is easily handled using the \textit{normal ordering }procedure. Unfortunately
this simple procedure is not applicable in curved spacetime (mainly
due to the non-existence of a unique time slicing). 

In 1965 DeWitt developed a regularization method \cite{Dewitt - Dynamical theory of groups and fields}
for such divergent quantities named \textit{point-splitting }or \textit{covariant
point separation}. DeWitt first illustrated the method for the regularization
of $\left\langle \phi^{2}\right\rangle $, and Christensen \cite{Christiansen}
later extended it to the stress-energy tensor. A key ingredient in
this method is separating the evaluation point into a pair of nearby
points $x,x'$ and then taking the coincidence limit $x'\to x$ while
subtracting some counter-term. This operation would presumably be
feasible if the (modes of the) field $\phi\left(x\right)$ were known
analytically. However, in BH backgrounds the field's modes need to
be computed numerically, and in such a case it becomes tremendously
difficult to implement the above limiting procedure, at least in the
direct naive way.

In the following years Candelas, Howard, Anderson and others developed
procedures aimed for implementing the point-splitting method numerically,
provided that one can compute the WKB approximation for the fields'
modes up to a sufficiently high order \cite{Candelas =000026 Howard - 1984 - phi2 Schwrazschild,Howard - 1984 - Tab Schwarzschild,Anderson - 1990 - phi2 static spherically symmetric,Anderson - 1995 - Tab static spherically symmetric}.
Alas, such computation of the WKB approximation is extremely difficult
for a generic background. Even in the spherically-symmetric static
case, the presence of a turning point makes the WKB expansion beyond
leading order a very hard task \textemdash{} let alone the case of
time-dependent background. For this reason most of these analyses
were carried in the euclidian sector (which, however, is usually restricted
to static situations). The most general case that was computed till
recently was the spherically-symmetric static background, by Anderson
\cite{Anderson - 1990 - phi2 static spherically symmetric} (for $\left\langle \phi^{2}\right\rangle $;
and later on this computation was extended by Anderson, Hiscock and
Samuel \cite{Anderson - 1995 - Tab static spherically symmetric}
to the renormalized stress-energy tensor). \cite{Ottewill - 2008 Kerr with a mirror} 

Recently we have developed a new approach for implementing the point-splitting
procedure numerically, which does not rely on the WKB approximation
at all (and can therefore be implemented directly in the Lorentzian
sector). Instead, it only requires that the background admits some
symmetry, which would allow mode decomposition of the field equation
(like the spherical-harmonic or $e^{-i\omega t}$ decompositions in
spherical or stationary backgrounds, respectively). The main idea
behind our method is that, since the point-splitting counter term
is known, one can decompose it and hence obtain from it ``mode-wise''
counter-terms that can be subtracted from the mode contributions,
thereby regularizing their sum (or integral). 

We developed several variants of this general method, which rely on
different possible symmetries of the backgrounds in consideration.
In the first paper \cite{Levi =000026 Ori - 2015 - t splitting regularization}
we have introduced the $t$-splitting variant which can be used in
stationary backgrounds. To simplify things we have chosen (as usual)
to first focus on the regularization of $\left\langle \phi^{2}\right\rangle $
rather than $\left\langle T_{\alpha\beta}\right\rangle $, as this
quantity is less divergent and also it is a scalar, which significantly
simplifies its presentation. 

Even though the $t$-splitting variant is very efficient, as it can
be used for every stationary background (e.g. a Kerr BH), it cannot
be used to study dynamical processes, which is the most interesting
case for us. In this paper we introduce the \textit{angular-splitting
}variant, which requires only spherical symmetry, and can be used
in dynamical backgrounds. This paper, too, will focus on $\left\langle \phi^{2}\right\rangle $
for simplicity. In two forthcoming papers we shall present the extension
of both the angular-splitting and $t$-splitting variants to the calculation
of the renormalized stress-energy tensor. 

The angular-splitting (or ``$\theta$ splitting'') variant is a
bit more complicated than $t$-splitting. In principle, in this variant
we aim to split the points in the $\theta$ direction, exploiting
the spherical-harmonics decomposition. It turns out, however, that
if the split is strictly in the angular direction we face an additional
divergence in an intermediate stage (integration over the frequency
$\omega$). In order to cure this intermediate divergence we have
to make an additional, smaller split \footnote{By ``smaller split'' we mean that we take this $t$-split to zero
\emph{before} taking the angular split to zero.} in the $t$ direction (see Sec. \ref{sec: The-angular-splitting-variant}).
This slightly complicates the regularization procedure, but it's worth
it, because the resultant method is a very powerful one, being applicable
to spherical dynamical backgrounds. In particular,  it should be
applicable  to evaporating spherical BHs. 

This paper is organized as follows: Section \ref{sec: Basic point-splitting}
briefly outlines the point-splitting procedure. In Sec. \ref{sec: The-angular-splitting-variant}
we present the angular-splitting method for spherically-symmetric
backgrounds, first in the static case and then also in the general
time-dependent case. Section \ref{sec: Schwarzschild} demonstrates
the application of this method to the Schwarzschild case. We give
the results for $\left\langle \phi^{2}\right\rangle _{ren}$ for a
massless scalar field in the various vacuum states (Boulware, Unruh,
and Hartle-Hawking). These results are compared with previous ones
to find excellent agreement. Finally, in Sec. \ref{sec: Discussion}
we summarize and discuss our results. 

\section{Field decomposition and basic point-splitting procedure\label{sec: Basic point-splitting}}

\subsection{Preliminaries}

The angular splitting is designed to allow regularization in asymptotically
flat, spherically-symmetric backgrounds, including time-dependent
ones. We thus consider here the general double-null spherically-symmetric
line element
\begin{equation}
ds^{2}=-\Gamma(u,v)dudv+r^{2}(u,v)d\Omega^{2}\label{eq:double-null metric}
\end{equation}
(with $\Gamma>0$), where $d\Omega^{2}=d\theta^{2}+\sin^{2}\theta d\varphi^{2}$.
It is also useful to define the corresponding space and time coordinates
$(t,z)$ via 
\[
v=t+z,\,u=t-z,
\]
in which the metric takes the form
\begin{equation}
ds^{2}=\Gamma\left(t,z\right)\left(-dt^{2}+dz^{2}\right)+r^{2}\left(t,z\right)d\Omega^{2}.\label{eq: Basic PS - Metric}
\end{equation}
By virtue of asymptotic flatness, we set $\Gamma\to1$ at spacelike
infinity $\left(z\to\infty\right)$.

In principle the coordinates $u,v$ are subject to the gauge freedom
$u\rightarrow u'(u),v\rightarrow v'(v)$. Employing asymptotic flatness
we choose $v$ to be an affine parameter along past null infinity
(PNI). We leave the gauge of $u$ unspecified for the time being,
except that we assume that $u\rightarrow-\infty$ at PNI and (in case
of a BH) $u\rightarrow\infty$ at the event horizon. In the static
case, however, $u$ is uniquely defined by the requirement of time-independent
metric.

Owing to spherical symmetry of the background metric we can decompose
the field in spherical harmonics $Y_{lm}$: 
\begin{equation}
\phi\left(x\right)=\sum_{lm}c_{lm}Y_{lm}\left(\theta,\varphi\right)\Psi_{l}\left(t,z\right)/r,\label{eq: spherical decompose}
\end{equation}
where hereafter $x$ denotes a spacetime point, and $c_{lm}$ are
arbitrary constants. The functions $\Psi_{l}\left(t,z\right)$ then
satisfy the 2D field equation 
\begin{equation}
\Psi_{l,uv}=-\frac{1}{4}V_{l}\,\Psi_{l},\label{eq: Basic 1+1 field eq - null}
\end{equation}
with the effective potential
\begin{equation}
V_{l}\left(u,v\right)=-4\frac{r_{,uv}}{r}+\Gamma\left[\frac{l\left(l+1\right)}{r^{2}}+m^{2}+\xi R\right],\label{eq:potential}
\end{equation}
where $R$ is the Ricci scalar. For later convenience we also write
the field equation using the $t,z$ coordinates
\[
\Psi_{l}''-\ddot{\Psi}_{l}=V_{l}\,\Psi_{l},
\]
where henceforth dot and prime denote derivatives with respect to
$t$ and $z$ respectively. In these coordinates the potential takes
the form
\begin{equation}
V_{l}\left(t,z\right)=\frac{r''-\ddot{r}}{r}+\Gamma\left[\frac{l\left(l+1\right)}{r^{2}}+m^{2}+\xi R\right].\label{eq:potential-1}
\end{equation}

\paragraph*{The $\omega$ modes: }

From this point on we shall restrict our attention to the massless
case $m=0$. Owing to asymptotic flatness $V_{l}$ vanishes at large
$r$, hence at PNI $\Psi_{l}$ asymptotically approaches some function
of $v$, which we may denote as $\Psi_{l}^{\infty}(v)$. \footnote{\label{fn: PNI data} To be more precise, the large-$r$ asymptotic
behavior of $\Psi_{l}$ takes the form $\Psi_{l}^{\infty}(v)+\zeta_{\omega l}\left(u\right)$,
where $\zeta_{\omega l}$ is some function of $u$. We refer to these
two terms as the data at PNI and FNI respectively.} 

For concreteness we shall assume at this stage that the spacetime
has a regular center (the case of eternal BH will be addressed later
on). Then the initial function $\Psi_{l}^{\infty}(v)$ \textemdash{}
along with the field equation (\ref{eq: Basic 1+1 field eq - null})
and the regularity condition 
\begin{equation}
\Psi_{l}\left(t,r=0\right)=0\label{eq: Basic PS - second boundary cond}
\end{equation}
at the center \textemdash{} uniquely determines the evolving solution
$\Psi_{l}\left(t,z\right)$. The harmonic initial functions $\Psi_{l}^{\infty}(v)=e^{-i\omega v}$
play a key role in the theory, and we shall denote the functions $\Psi_{l}\left(t,z\right)$
which evolve from such harmonic PNI initial data by $\Psi_{\omega l}\left(t,z\right)$.
Note that these functions depend on both $\omega$ (through initial
conditions) and $t$. Using these $\omega l$ mode functions we can
decompose the field as 
\begin{equation}
\phi\left(x\right)=\sum_{lm}\int_{0}^{\infty}d\omega\,c_{\omega lm}\,f_{\omega lm}\left(x\right),\label{eq: Basic field decomposition}
\end{equation}
where $c_{\omega lm}$ are arbitrary expansion coefficients and
\begin{equation}
f_{\omega lm}\left(x\right)=Y_{lm}\left(\theta,\varphi\right)\bar{\Psi}_{\omega l}\left(t,z\right),\label{classical field decompose}
\end{equation}
where
\begin{equation}
\bar{\Psi}_{\omega l}\left(t,z\right)\equiv\frac{\Psi_{\omega l}\left(t,z\right)}{r\sqrt{4\pi\omega}}\,.\label{classical field decompose-1}
\end{equation}
The factor $1/\sqrt{4\pi\omega}$ was inserted in order for the $f_{\omega lm}$
modes to be properly Klein-Gordon normalized. Note that the $f_{\omega lm}$
functions satisfy the basic field equation (\ref{eq: Intro - KG eq.})
as well as the decomposed equation (\ref{eq: Basic 1+1 field eq - null}).

\subsection{Quantum field:}

The quantum field operator is constructed from the $f_{\omega lm}$
mode functions: 
\begin{equation}
\phi\left(x\right)=\sum_{l=0}^{\infty}\int_{0}^{\infty}d\omega\sum_{m=-l}^{l}\left(f_{\omega lm}\left(x\right)a_{\omega lm}+f_{\omega lm}^{*}\left(x\right)a_{\omega lm}^{\dagger}\right),\label{eq: Basic PS - field expansion}
\end{equation}
where $a_{\omega lm},a_{\omega lm}^{\dagger}$ are the creation and
annihilation operators. We point out that one can choose different
orders for the summation/integration operations. Here we choose the
order which best suits our regularization procedure: Since we split
in $\theta$, the associated operation of summation over $l$ should
better be the last one. The order of $\omega$-integration and $m$-summation
is less crucial, however the one selected here is more convenient.
Note that the field decomposition (\ref{eq: Basic PS - field expansion})
naturally defines the \emph{vacuum state} $\left|0\right\rangle $,
which is annihilated by every $a_{\omega lm}$, namely $a_{\omega lm}\left|0\right\rangle =0$
for every $\omega lm$.

In the case of an eternal BH there is no regular center, instead there
is a past horizon. One then has to introduce another set of modes
defined with their own boundary conditions. This is addressed in Sec.
\ref{subsec: Two-ended static} and \ref{subsec: Two-ended dynamic}
(for the static and time-dependent cases, respectively). 

\subsection{Calculation of $\left\langle \phi^{2}\right\rangle _{ren}$\label{subsec:Calculation-of-:}}

Trying to naively calculate the vacuum expectation value of $\phi^{2}$
yields a divergent mode-sum
\begin{equation}
\left\langle \phi^{2}\left(x\right)\right\rangle _{naive}=\hbar\sum_{l=0}^{\infty}\int_{0}^{\infty}d\omega\sum_{m=-l}^{l}\left|Y_{lm}\left(\theta,\varphi\right)\right|^{2}\left|\bar{\Psi}_{\omega l}\left(t,z\right)\right|^{2}.\label{eq: Basic PS - naive phi2}
\end{equation}
As mentioned above one can consider different orders of summation
and integration, yet they are all divergent. In the specific ordering
(\ref{eq: Basic PS - naive phi2}) the sum over $m$ converges of
course (it is a finite sum), but the integral over $\omega$ diverges
logarithmically. Furthermore, even after the divergence of the integral
over $\omega$ is cured (as described below), the sum over $l$ is
also divergent, and even more strongly (like $l^{2}\ln l$). 

In the calculation of various renormalized quantities one often faces
the situation in which an integral over $\omega$ fails to converge
due to oscillations of the integrand at large $\omega$. A similar
problem of oscillations may also be faced in the summation over $l$
(usually when the two points are separated in $\theta$). To handle
such non-convergent oscillations we employ the concept of \emph{generalized
integral} (or \emph{generalized sum}), in which the oscillations are
damped by multiplying the integrand (or sequence) by some factor exponentially-decaying
in $\omega$ (or $l$) \textemdash{} and subsequently taking the limit
of vanishing exponential pre-factor. This is described in more detail
in Appendix \ref{sec: Generalized-sums}. We have already faced this
problem of non-converging oscillatory integral over $\omega$ in our
$t$-splitting variant. In Ref. \cite{Levi =000026 Ori - 2015 - t splitting regularization}
we explained the geometric origin of these large-$\omega$ oscillations
(due to connecting null geodesics), and described their curing by
generalized integration. We also prescribed our pragmatic method for
implementing the generalized integral by \emph{self-cancellation}
of the oscillations. It is important to emphasize that all the sums
and integrals in this paper are (at least in principle \footnote{\label{fn: convensional sums} Some of the integrals/sums in this
paper do converge in the usual sense, but  we are still allowed to
regard them as generalized ones, for the following obvious reason:
Whenever an integral/sum converges in the usual sense, it is guaranteed
to coincide with the corresponding generalized integral/sum (see Appendix
\ref{sec: Generalized-sums}). }) \emph{generalized} ones. Thus, by stating that the mode-sum in Eq.
(\ref{eq: Basic PS - naive phi2}) diverges we actually mean it diverges
even when the integral and sum are generalized. 

\paragraph*{Point splitting:}

In 1965 DeWitt showed \cite{Dewitt - Dynamical theory of groups and fields}
that one can consider the two-point function $\left\langle \phi\left(x\right)\phi\left(x'\right)\right\rangle $,
and obtain a meaningful (renormalized) expectation value of $\phi^{2}$
by taking the coincidence limit
\begin{equation}
\left\langle \phi^{2}\left(x\right)\right\rangle _{ren}=\lim_{x'\to x}\left[\left\langle \phi\left(x\right)\phi\left(x'\right)\right\rangle -G_{DS}\left(x,x'\right)\right].\label{eq: Basic PS - phi ren limit}
\end{equation}
Here $G_{DS}\left(x,x'\right)$ is the DeWitt-Schwinger \textit{counter-term,}\textit{\emph{
a locally-constructed quantity which captures the singular piece of
the }}two-point function\textit{\emph{ at the limit $x'\to x$.}}
For a scalar field with mass $m$ \footnote{\label{fn: massive} Although this paper mainly addresses the massless
case, for completeness we treat here the counter-term also in the
$m\neq0$ case.} and coupling constant $\xi$ it is\textit{ 
\begin{equation}
\frac{1}{\hbar}G_{DS}\left(x,x'\right)=\frac{1}{8\pi^{2}\sigma}+\frac{m^{2}+\left(\xi-1/6\right)R}{8\pi^{2}}\left[\gamma+\frac{1}{2}\ln\left(\frac{\mu^{2}\sigma}{2}\right)\right]-\frac{m^{2}}{16\pi^{2}}+\frac{1}{96\pi^{2}}R_{\alpha\beta}\frac{\sigma^{;\alpha}\sigma^{;\beta}}{\sigma},\label{eq: Basic PS - Counter term}
\end{equation}
}where $R_{\alpha\beta},R$ are the Ricci tensor and Ricci scalar,
$\gamma$ is Euler constant; and $\sigma$ is the bi-scalar of the
short geodesic connecting $x$ to $x'$, which is equal to half the
geodesic distance squared (see Ref. \cite{Dewitt - Dynamical theory of groups and fields}).
The quantity $\mu$ is an unknown parameter, representing the well
known ambiguity in the regularization process. 

In the few cases that the modes are known analytically the recipe
given by DeWitt can presumably be directly used to calculate $\left\langle \phi^{2}\left(x\right)\right\rangle _{ren}$,
e.g. in the case of Robertson\textendash Walker background \cite{Bunch =000026 Devies}.
However in most cases of interest, and particularly for BH backgrounds,
the mode functions are known only numerically and with limited accuracy;
namely $\bar{\Psi}_{\omega l}\left(t,z\right)$ is computed numerically
for some finite range in $\omega$, from zero to some $\omega_{max}$
(and for some range of $l\le l_{max}$). Evaluation of the coincidence
limit in Eq. (\ref{eq: Basic PS - phi ren limit}) then becomes an
extremely difficult task. As $x'\to x$ the $\omega_{max}$ and $l_{max}$
values required for effective convergence grow rapidly, typically
like the inverse of the separation. 

Our approach of mode-sum regularization is tailored to overcome this
difficulty: Essentially we handle the coincidence limit \emph{analytically},
translating it to a certain regularization process which we implement
while summing/integrating over the modes. Namely, we subtract certain
functions of $\omega$ and $l$ upon summation/integration. The entire
numerical part of the calculation \textemdash{} the evaluation of
the mode contributions and their sum/integral \textemdash{} is actually
done \emph{at coincidence}, which makes the entire numerical scheme
tractable. We already described the application of this approach for
$t$-splitting in Ref. \cite{Levi =000026 Ori - 2015 - t splitting regularization}.
Here we shall describe its application to $\theta$-splitting. 

\section{The angular-splitting variant\label{sec: The-angular-splitting-variant}}

\subsection{The static case\label{subsec: The-stationary-case}}

In order to make the regularization method more transparent we first
present it for the special case of a static metric with a regular
center (namely no eternal BH). In subsection \ref{subsec: Two-ended static}
we describe the adjustment needed for an eternal BH background, and
in Sec. \ref{subsec: The-general-spherically-symmetric-case} we generalize
it for dynamical backgrounds.

In the static case the general spherically-symmetric line element
is 
\begin{equation}
ds^{2}=\Gamma\left(z\right)\left(-dt^{2}+dz^{2}\right)+r^{2}(z)d\Omega^{2}.\label{eq: Stationary case - Metric}
\end{equation}
The field is decomposed as in Eq. (\ref{classical field decompose}),
but owing to time-translation symmetry the $t$ dependence of $\bar{\Psi}_{\omega l}$
is now trivial: 
\[
\bar{\Psi}_{\omega l}\left(t,z\right)=e^{-i\omega t}\,\bar{\psi}_{\omega l}(z)\,,
\]
hence the mode decomposition becomes 
\[
f_{\omega lm}\left(x\right)=e^{-i\omega t}\,Y_{lm}\left(\theta,\varphi\right)\bar{\psi}_{\omega l}\left(z\right).
\]
We again introduce the auxiliary function
\begin{equation}
\psi_{\omega l}\left(z\right)=r\,\sqrt{4\pi\omega}\,\bar{\psi}_{\omega l}\left(z\right)\label{eq:auxiliary}
\end{equation}
which obeys the simple one-dimensional ODE 
\begin{equation}
\psi''_{\omega l}=\left[V_{l}\left(z\right)-\omega^{2}\right]\psi_{\omega l}\,,\label{eq: Stationary case - field eq-1}
\end{equation}
and the potential (\ref{eq:potential-1}) now reduces to 
\begin{equation}
V_{l}\left(z\right)=\frac{r''}{r}+\Gamma\left[\frac{l\left(l+1\right)}{r^{2}}+\xi R\right].\label{eq: Stationary case - Potential-1}
\end{equation}
 (Recall that we restrict the analysis to massless fields.)

The boundary conditions for $\psi_{\omega l}$ are set such that the
incoming monochromatic wave has a unit amplitude, and the modes are
regular at the center. Owing to the presence of regular center, the
reflected wave must have the same amplitude as the incoming one. \footnote{This follows from the fact that at $r\to0$ there exist, for any $l$,
a regular solution ($\psi\propto r^{l+1}$) and a singular solution
($\psi\propto r^{-l}$). Since the radial equation is real, it immediately
follows that the regular solution must be essentially real (namely,
real, up to a constant pre-factor; otherwise, there would exist a
second, independent, regular solution $\psi^{*}$). If the amplitudes
of the outgoing and ingoing waves were different, the overall large-$r$
asymptotic solution would fail to be (essentially) real.} Thus, the boundary conditions take the form
\begin{equation}
\psi_{\omega l}\left(r=0\right)=0\,\,,\,\,\,\,\,\lim_{z\to\infty}\psi_{\omega l}\left(z\right)=e^{-i\omega z}+e^{i\lambda\left(\omega,l\right)}e^{i\omega z}\,,\label{eq: Stationary case - Boundary cond.}
\end{equation}
where $\lambda\left(\omega,l\right)$ is an unknown (real) phase associated
to the reflected modes.

\subsubsection{The integral over $\omega$ \label{subsec:The-integral-over}}

The main essence of our regularization method is to split the points
in a direction of symmetry, and to choose the order of the mode-sum
operations such that the sum (or integral) that corresponds to the
splitting direction is the last to be performed. Correspondingly,
in the $\theta$-splitting variant we preform the sum over $l$ last.
Thus, naively we would like to implement the point-splitting procedure
in the following manner:
\[
\left\langle \phi^{2}\left(x\right)\right\rangle _{split(naive)}=\lim_{\varepsilon\to0}\left[\hbar\sum_{l=0}^{\infty}\int_{0}^{\infty}d\omega\sum_{m=-l}^{l}Y_{lm}\left(\theta,\varphi\right)Y_{lm}^{*}\left(\theta+\varepsilon,\varphi\right)\left|\bar{\psi}_{\omega l}\left(z\right)\right|^{2}-G_{DS}\left(x,x'\right)\right].
\]
The sum over $m$ is straightforward, 
\begin{equation}
\sum_{m=-l}^{l}Y_{lm}\left(\theta,\varphi\right)Y_{lm}^{*}\left(\theta+\varepsilon,\varphi\right)=\frac{2l+1}{4\pi}P_{l}\left(cos\,\varepsilon\right),\label{eq: m-sum}
\end{equation}
hence the first term in the squared brackets becomes 
\[
\hbar\sum_{l=0}^{\infty}\frac{2l+1}{4\pi}P_{l}\left(cos\,\varepsilon\right)\int_{0}^{\infty}\left|\bar{\psi}_{\omega l}\left(z\right)\right|^{2}d\omega.
\]
It turns out, however, that the integral over $\omega$ still diverges.
In fact, this integral is not affected at all by the splitting in
$\theta$, as one can easily verify. \footnote{Unlike what one might naively expect, certain mode-sum operations
may diverge even when the points are separated. The convergence or
otherwise of the mode-sum operations may depend on the splitting direction
as well as on the order of these operations.} To see this divergence explicitly one can examine the large-$\omega$
asymptotic behavior of the modes. This large-$\omega$ analysis is
presented in Appendix \ref{sec: Large w approximation} (with fairly
detailed description of the analysis in the case of static eternal
BH, and summary of final results for the other, more complicated,
cases). For a background with a regular center we obtain (see Sec.
\ref{subsec:Regular-center}) 
\begin{equation}
\left|\bar{\psi}_{\omega l}\left(z\right)\right|^{2}=\frac{1}{2\pi r^{2}\omega}+(...)\,,\label{eq: Omega integral - psi2 leading order}
\end{equation}
where ``$(...)$'' denotes terms whose integral over $\omega$ converge.
(Specifically it includes $\propto\omega^{-3}$ terms and also purely
oscillatory terms whose amplitude decays as $1/\omega$.) Hence its
integral over $\omega$ diverges logarithmically.

To overcome this divergence we introduce an additional splitting in
the $t$ direction, namely 
\[
x=(t,z,\theta,\varphi),\,x'=(t+\delta,z,\theta+\varepsilon,\varphi).
\]
However, this split in $t$ is taken to be ``small'', in the following
sense: When implementing the coincidence limit we first take the limit
$\delta\to0$ and only afterwards the limit $\varepsilon\to0$. Recalling
Eq. (\ref{eq: m-sum}), the renormalized vacuum expectation value
of $\phi^{2}$ now takes the form 
\[
\left\langle \phi^{2}\left(x\right)\right\rangle _{ren}=\lim_{\varepsilon\to0}\,\lim_{\delta\to0}\left[\hbar\sum_{l=0}^{\infty}\int_{0}^{\infty}d\omega\,\frac{2l+1}{4\pi}\,P_{l}\left(cos\,\varepsilon\right)\left|\bar{\psi}_{\omega l}\left(z\right)\right|^{2}e^{i\omega\delta}-G_{DS}\left(x,x'\right)\right].
\]
Since $G_{DS}\left(x,x'\right)$ is regular even for $\delta=0$ (as
long as $\varepsilon>0$), it is possible to rewrite this expression
as 
\begin{equation}
\left\langle \phi^{2}\left(x\right)\right\rangle _{ren}=\lim_{\varepsilon\to0}\left[\lim_{\delta\to0}\hbar\sum_{l=0}^{\infty}\,\frac{2l+1}{4\pi}\,P_{l}\left(cos\,\varepsilon\right)\int_{0}^{\infty}\left|\bar{\psi}_{\omega l}\left(z\right)\right|^{2}e^{i\omega\delta}d\omega-G_{DS}\left(\varepsilon\right)\right].\label{eq: Omega integral - phi2 two limits}
\end{equation}
Note the role of the two limits in this expression: The limit $\delta\to0$
regulates the $\omega$-integral, and subsequently the limit $\varepsilon\to0$
controls the sum over $l$.

The counter-term $G_{DS}$ is expressed in Eq. (\ref{eq: Basic PS - Counter term})
as a function of the geodesic bi-scalar $\sigma$. Our calculation
scheme requires us to re-express $G_{DS}$ in terms of $\varepsilon$.
In fact we find it most useful to expand $G_{DS}$ in powers of $\sin(\varepsilon/2)$.
\footnote{This makes the Legendre decomposition (used below) simpler. Also in
the Minkowski case, which provides a very useful guide, $\sigma$
and $G_{DS}$ are \emph{exactly} proportional to $\sin^{2}(\varepsilon/2)$
and $\sin^{-2}(\varepsilon/2)$ respectively.} To this end, in the first stage we expand the geodesic equation (integrated
from $\theta$ to $\theta+\varepsilon$) to obtain $\sigma$ Taylor-expanded
in $\varepsilon$. This expansion is rather lengthy but is nevertheless
straightforward, and it can be automated using standard algebraic-computation
software. It yields 
\begin{equation}
\sigma=\frac{r^{2}}{2}\,\varepsilon^{2}+\tilde{c}(z)\varepsilon^{4}+O(\varepsilon^{5})\,\,,\label{eq: Omega integral - sigma}
\end{equation}
where
\[
\tilde{c}\left(z\right)=-\frac{r^{2}}{24\Gamma}\,r'^{2}\,.
\]
Then we re-expand it in $\sin(\varepsilon/2)$, and substitute in
Eq. (\ref{eq: Basic PS - Counter term}). We obtain the counter-term
in the form
\begin{equation}
\frac{1}{\hbar}G_{DS}\left(\varepsilon\right)=a\left(z\right)\sin^{-2}\left(\varepsilon/2\right)+c\left(z\right)\left[\ln\left(\mu\,r\sin\left(\varepsilon/2\right)\right)+\gamma\right]+d\left(z\right)+O(\varepsilon)\,,\label{eq: Omega integral - Gds(epsilon)}
\end{equation}
where $a\left(z\right),c\left(z\right),d\left(z\right)$ are coefficients
that (in the massless case) take the form 
\[
a\left(z\right)=\frac{1}{16\pi^{2}r^{2}}\,\,,
\]
\[
c\left(z\right)=\frac{\left(1/6-\xi\right)}{8\pi^{2}r^{2}\Gamma^{3}}\left[-2\Gamma^{3}-r^{2}\Gamma'^{2}+r^{2}\Gamma\Gamma''+2\Gamma^{2}r'^{2}+4r\Gamma^{2}r''\right]\,\,,
\]
\[
d\left(z\right)=-\frac{1}{48\pi^{2}r\Gamma}\,r''\,\,.
\]

The integral in Eq. (\ref{eq: Omega integral - phi2 two limits})
is regularized by the oscillating factor $e^{i\omega\delta}$, which
provides an effective cutoff at $\omega\sim1/\delta$. In order to
regularize this integral at the $\delta\to0$ limit, we next process
this integral by the usual technique of adding and subtracting some
$\omega$-dependent quantity, which depicts the large-$\omega$ leading
order of the integrand. (A successful subtraction of the divergent
piece will eventually enable us to take the limit $\delta\to0$ already
in the integrand.) Utilizing the fact that the divergent piece in
Eq. (\ref{eq: Omega integral - psi2 leading order}) is independent
of $l$, we found that a convenient way to regularize the $\delta\to0$
limit is to add and subtract the corresponding contribution from the
$l=0$ mode. Namely we write the integral in Eq. (\ref{eq: Omega integral - phi2 two limits})
as the sum of two integrals: 
\begin{equation}
\int_{0}^{\infty}E_{\omega l}\,e^{i\omega\delta}d\omega=\int_{0}^{\infty}\left[E_{\omega l}-E_{\omega,l=0}\right]e^{i\omega\delta}d\omega+\int_{0}^{\infty}E_{\omega,l=0}\,e^{i\omega\delta}d\omega\,,\label{eq: Omega integral - break to two integrals}
\end{equation}
where for brevity we have defined 
\begin{equation}
E_{\omega l}\left(z\right)\equiv\left|\bar{\psi}_{\omega l}\left(z\right)\right|^{2},\label{eq: E omega l defined}
\end{equation}
which represents the integrand at the coincide $\left(\delta=0\right)$.
The first integral in the right-hand side now converges even for $\delta=0$,
so it is possible to take the limit by setting $\delta=0$ already
in the integrand. \footnote{\label{fn: Interchange} This involves interchanging the order of
the $\delta\to0$ limit with the sum over $l$ and the integral over
$\omega$. Although the mathematical justification of such an interchange
is far from obvious, we find strong evidence for its validity from
the asymptotic behavior of the various quantities at large $\omega$
and $l$. Other interchanges of operations are also used in a few
other occasions later on. This issue is further discussed in Sec.
\ref{sec: Discussion}.} For briefness we denote this integral (with the limit $\delta\to0$
already taken) by
\begin{equation}
F\left(l\right)\equiv\int_{0}^{\infty}\left[E_{\omega l}-E_{\omega,l=0}\right]d\omega.\label{eq: Omega integral - Define Fl}
\end{equation}
This quantity can be computed directly once the modes $\bar{\psi}_{\omega l}\left(r\right)$
are known (even numerically). Note that $F(l)$ (like $E_{\omega l}$
and $G_{DS}$, and like $Z$ below) depends on $z$ as well. We often
omit this $z$-dependence for brevity, although below we occasionally
denote it as $F\left(l,z\right)$ when appropriate. 

The second integral in Eq. (\ref{eq: Omega integral - break to two integrals})
is more interesting, and we denote it as 
\[
Z\left(\delta\right)\equiv\int_{0}^{\infty}E_{\omega,l=0}e^{i\omega\delta}d\omega\,.
\]
It is a well defined function of $\delta$ that diverges as $\delta\to0$
(but finite otherwise). Note that $Z\left(\delta\right)$ also depends
on $z$, but it does \emph{not} depend on $l$. 

Substituting the $\omega$-integral in (\ref{eq: Omega integral - break to two integrals})
{[}which is now represented by $F\left(l\right)+Z\left(\delta\right)${]}
back in Eq. (\ref{eq: Omega integral - phi2 two limits}) yields
\begin{equation}
\left\langle \phi^{2}(x)\right\rangle _{ren}=\lim_{\varepsilon\to0}\left[\hbar\sum_{l=0}^{\infty}\frac{2l+1}{4\pi}P_{l}\left(cos\,\varepsilon\right)F\left(l\right)+\hbar\lim_{\delta\to0}\left\{ \sum_{l=0}^{\infty}\frac{2l+1}{4\pi}P_{l}\left(cos\,\varepsilon\right)Z\left(\delta\right)\right\} -G_{DS}\left(\varepsilon\right)\right].\label{eq: Omegan integral - phi2 with the null term}
\end{equation}
A crucial observation is that for any (small) \footnote{\label{fn: condition} To be more precise, the condition for this
vanishing of the $l$-sum is $\varepsilon\neq n\pi$ for any integer
$n$.} finite $\varepsilon$ the term in curly brackets actually \emph{vanishes},
because 
\[
\sum_{l=0}^{\infty}(2l+1)P_{l}\left(cos\,\varepsilon\right)=0.
\]
This is shown in Appendix \ref{sec: Legendre-sums}. Equation (\ref{eq: Omegan integral - phi2 with the null term})
thus takes the simpler form
\begin{equation}
\left\langle \phi^{2}\left(x\right)\right\rangle _{ren}=\lim_{\varepsilon\to0}\left[\hbar\sum_{l=0}^{\infty}\frac{2l+1}{4\pi}P_{l}\left(cos\,\varepsilon\right)F\left(l\right)-G_{DS}\left(\varepsilon\right)\right].\label{eq: Omega integral - ph2 final res}
\end{equation}

In summary, we have been able to take the limit $\delta\to$0, namely
to carry the first step of the regularization. This leaves us with
the limit $\varepsilon\to0$ and the sum over $l$, which we next
consider.

\subsubsection{The sum over $l$}

The situation in Eq. (\ref{eq: Omega integral - ph2 final res}) is
in some respect similar to the $t$-splitting variant \cite{Levi =000026 Ori - 2015 - t splitting regularization}
(see in particular Eq. (3.5) therein). The main difference is that
the $\omega$-integral of \cite{Levi =000026 Ori - 2015 - t splitting regularization}
is here replaced by sum over $l$. This directly reflects on the way
we treat the counter-term: The Fourier decomposition of $G_{DS}\left(\varepsilon\right)$
in Ref. \cite{Levi =000026 Ori - 2015 - t splitting regularization}
will be replaced here by a Legendre decomposition. From Eq. (\ref{eq: Omega integral - Gds(epsilon)}),
we see that the quantities that need be Legendre-expanded are $\sin^{-2}\left(\varepsilon/2\right)$
and $\ln[\sin\left(\varepsilon/2\right)]$. This decomposition yields
(see Appendix \ref{sec: Legendre-decomposition}) 
\begin{equation}
\sin^{-2}\left(\varepsilon/2\right)=-8\pi\sum_{l=0}^{\infty}\frac{2l+1}{4\pi}h\left(l\right)P_{l}\left(cos\,\varepsilon\right)\label{eq:sin}
\end{equation}
and 

\begin{equation}
\ln[\sin\left(\varepsilon/2\right)]=2\pi\sum_{l=0}^{\infty}\frac{2l+1}{4\pi}\Lambda\left(l\right)P_{l}\left(cos\,\varepsilon\right)\,,\label{eq:LogSin}
\end{equation}
where $h\left(l\right)$ is the \textit{Harmonic Number }given by
\[
h\left(l\right)\equiv\sum_{k=1}^{l}\frac{1}{k}\;,
\]
(with $h\left(0\right)\equiv0$), and $\Lambda\left(l\right)$ is
defined to be
\[
\Lambda\left(l\right)\equiv\begin{cases}
-1 & l=0\\
-\frac{1}{l\left(l+1\right)} & l>0
\end{cases}\;.
\]
Inserting these identities in Eqs. (\ref{eq: Omega integral - Gds(epsilon)})
and (\ref{eq: Omega integral - ph2 final res}) results in
\begin{equation}
\left\langle \phi^{2}\left(x\right)\right\rangle _{ren}=\hbar\lim_{\varepsilon\to0}\sum_{l=0}^{\infty}\frac{2l+1}{4\pi}P_{l}\left(cos\,\varepsilon\right)F_{reg}\left(l,z\right)+W\left(z\right),\label{eq: l sum - phi2 ren}
\end{equation}
where
\begin{gather}
W\left(z\right)=-\hbar\left[\left(\ln\left(\mu\,r\right)+\gamma\right)c\left(z\right)+d\left(z\right)\right],\label{eq: l sum - W}
\end{gather}
\begin{equation}
F_{reg}\left(l,z\right)\equiv F\left(l,z\right)-F_{sing}\left(l,z\right),\label{eq: l sum - Freg}
\end{equation}
and
\begin{equation}
F_{sing}\left(l,z\right)=-8\pi a\left(z\right)h\left(l\right)+2\pi c\left(z\right)\Lambda\left(l\right).\label{eq: l sum - Fsing}
\end{equation}
Note that $F_{sing}$ diverges logarithmically with $l$, like $h(l)$. 

\subsubsection{``Blind spots'' and their self-cancellation \label{subsec: Blind-spots}}

After we have subtracted the singular piece $F_{sing}\left(l,z\right)$,
one might hope that the sum over $l$ in Eq. (\ref{eq: l sum - phi2 ren})
would now converge even when $\varepsilon=0$ is substituted in the
Legendre polynomial. Unfortunately this is not the case, and one generally
finds this sum diverges if $P_{l}\rightarrow1$ is substituted. This
is demonstrated below in the Schwarzschild case (see e.g. Fig. \ref{fig: figure 2b}
in Sec. \ref{sec: Schwarzschild}). This divergence indicates that
the \textit{\emph{counter-term $G_{DS}$}}\textit{ }does not provide
full information about the mode-sum singularity (stated in other words,
we loose some of the information in the Legendre decomposition). To
understand this phenomenon, consider the sum $\sum_{l=0}^{\infty}\left(2l+1\right)$
which is obviously divergent, while for any (small) $\varepsilon>0$
\[
\sum_{l=0}^{\infty}(2l+1)P_{l}\left(cos\,\varepsilon\right)=0
\]
(see Appendix \ref{sec: Legendre-sums}; and recall however footnote
\ref{fn: condition}). We shall refer to this phenomenon as a \emph{blind
spot}. We define a ``blind spot'' as a function $B\left(l\right)$
for which for any (small) $\varepsilon\neq0$,
\[
\sum_{l=0}^{\infty}(2l+1)B\left(l\right)P_{l}\left(cos\,\varepsilon\right)=0,
\]
and yet $\sum_{l=0}^{\infty}\left(2l+1\right)B\left(l\right)$ diverges.
In Appendix \ref{sec: Legendre-sums} we show that all the functions
of the form 
\begin{equation}
B\left(l\right)=const\cdot\left[l\left(l+1\right)\right]^{n}\,;\,\,\,\,n=0,1,2,3,...\,,\label{eq: Blind-spots - general BS form}
\end{equation}
are blind spots. We shall assume that these are the only blind spots
that show up in $\theta$-splitting. We cannot prove this assumption,
but nevertheless this is the only type we encountered so far in angular
splitting (e.g. in Schwarzschild and Reissner-Nordstrom backgrounds,
including also in the calculation of $\left\langle T_{\alpha\beta}\right\rangle _{ren}$
in these spacetimes). Furthermore, even if one encounters a blind
spot of a different type in some background metric, it is reasonable
to assume that it will be possible to handle this new type as well. 

It is also reasonable to assume that the $\sin^{-2}\left(\varepsilon/2\right)$
term in the \textit{\emph{counter-term}} should account for the most
divergent part in $F\left(l,r\right)$. Then we can conclude that
the only possible blind spot in the regularization of $\phi^{2}$
corresponds to $n=0$, namely, $B\left(l\right)=const\equiv B_{0}(z)$.
We shell assume that this is indeed the case (but note that if blind
spots that correspond to other $n$ values happen to show up we know
how to self-cancel them as well). We can therefore write
\begin{equation}
F_{reg}\left(l,z\right)=B_{0}\left(z\right)+A\left(l,z\right),\label{eq: Blind-spots - Freg}
\end{equation}
where $\sum_{l=0}^{\infty}\left(2l+1\right)A\left(l,z\right)$ is
assumed to be convergent (numerically we find that $A\left(l,z\right)$
decays faster than $l^{-3}$). We can therefore replace $F_{reg}\left(l,z\right)$
by $A\left(l,z\right)$ in Eq. (\ref{eq: l sum - phi2 ren}). The
sum over $l$ in the R.H.S of the latter now converges \textemdash{}
even for $\varepsilon=0$. Presumably we can now interchange the limit
with the sum in Eq. (\ref{eq: l sum - phi2 ren}), which then reduces
to
\begin{equation}
\left\langle \phi^{2}\left(x\right)\right\rangle _{ren}=\hbar\sum_{l=0}^{\infty}\frac{2l+1}{4\pi}\,A\left(l,z\right)+W\left(z\right)\,.\label{eq: l sum - phi2 ren-1-2}
\end{equation}

We still need to address one pragmatic issue: The quantity we obtain
from the numerics (via Eq. (\ref{eq: l sum - Freg})) is $F_{reg}\left(l,z\right)$
rather than $A\left(l,z\right)$. We somehow need to subtract the
appropriate quantity $B_{0}\left(z\right)$ from $F_{reg}\left(l,z\right)$.
In principle we can do this by picking some sufficiently large $l$
value which we denote $l_{large}$ {[}such that $A\left(l,z\right)$
is sufficiently small and can be neglected{]}, and hence approximate
$B_{0}$(z) by $F_{reg}\left(l_{large},z\right)$; This way, we approximate
$A\left(l,z\right)$ by $F_{reg}\left(l,z\right)-F_{reg}\left(l_{large},z\right)$.
We call this strategy ``self-cancellation'' of the undesired quantity
$B_{0}$(z). \footnote{This is somewhat analogous to the process of oscillation self-cancellation
in the $t$-splitting \cite{Levi =000026 Ori - 2015 - t splitting regularization}.} 

We find it more convenient, however, to achieve the self cancellation
of $B_{0}$ in a slightly different way: We define the sequence of
partial sums:

\begin{equation}
H\left(l,z\right)\equiv\sum_{k=0}^{l}\frac{2k+1}{4\pi}\left[F_{reg}\left(k,z\right)-F_{reg}\left(l,z\right)\right].\label{eq: Blind-spots - H}
\end{equation}
The desired sum over $l$ in Eq. (\ref{eq: l sum - phi2 ren-1-2})
is nothing but the limit $l\to\infty$ of $H\left(l,z\right)$: 

\begin{multline*}
\lim_{l\to\infty}H\left(l,z\right)=\lim_{l\to\infty}\sum_{k=0}^{l}\frac{2k+1}{4\pi}\,\left[A\left(k,z\right)-A\left(l,z\right)\right]\\
=\sum_{k=0}^{\infty}\frac{2k+1}{4\pi}\,A\left(k,z\right)+\frac{1}{4\pi}\,\lim_{l\to\infty}\left[\left(l+1\right)^{2}\,A\left(l,z\right)\right].
\end{multline*}
In the last term the limit $l\to\infty$ vanishes, implying that 
\[
\lim_{l\to\infty}H\left(l,z\right)=\sum_{l=0}^{\infty}\frac{2l+1}{4\pi}\,A\left(l,z\right).
\]
We therefore rewrite Eq. (\ref{eq: l sum - phi2 ren-1-2}) in the
form 
\begin{equation}
\left\langle \phi^{2}\left(x\right)\right\rangle _{ren}=\hbar\lim_{l\to\infty}H\left(l,z\right)+W\left(z\right).\label{eq: Blind-spots - phi2 final}
\end{equation}

We have thus demonstrated how a self-cancellation process can turn
the generalized sum with separated points in Eq. (\ref{eq: l sum - phi2 ren})
into a convergent conventional sum/limit at the coincide, which can
be directly computed numerically. We have tailored our self-cancellation
procedure to the blind spot we have encountered so far (namely $n=0$).
The generalization to any (integer) $n\geq1$ is straightforward;
and if another type of blind spot shows up in some background, we
assume it would be possible to self-cancel it too. Furthermore, methods
other then self-cancellation can be used to get rid of the blind spots.
One such example is to calculate $B_{0}\left(z\right)$ from leading-order
WKB approximation, and subtract it instead of $F_{reg}(l,z)$ in Eq.
(\ref{eq: Blind-spots - H}).

Let us summarize the result of our regularization procedure: The final
expression for $\left\langle \phi^{2}\left(x\right)\right\rangle _{ren}$
is given in Eq. (\ref{eq: Blind-spots - phi2 final}), with $H\left(l,z\right)$
defined in Eq. (\ref{eq: Blind-spots - H}), where $F_{reg}\left(l,z\right)$
and $W\left(z\right)$ are specified in (\ref{eq: l sum - W}-\ref{eq: l sum - Freg}).
The quantity $F_{reg}\left(l,z\right)$ is defined using $F_{sing}\left(l,z\right)$
of Eq. (\ref{eq: l sum - Fsing}) and $F\left(l,z\right)$ which is
numerically computed according to Eqs. (\ref{eq: E omega l defined},\ref{eq: Omega integral - Define Fl})
from the mode functions $\bar{\psi}_{\omega l}\left(z\right)$.

\subsubsection{The static eternal BH case\label{subsec: Two-ended static}}

In the case of a static eternal BH $z$ runs from $\infty$ at spacelike
infinity to $-\infty$ at the horizon. In this case there are two
sets of basis solutions for the radial equation, instead of one. These
two sets, which we denote $\psi_{\omega l}^{in},\psi_{\omega l}^{up}$,
are defined by the boundary conditions 
\begin{gather}
r\,\sqrt{4\pi\omega}\,\bar{\psi}_{\omega l}^{in}(z)\equiv\psi_{\omega l}^{in}\left(z\right)=\begin{cases}
\tau_{\omega l}^{in}\,e^{-i\omega z}, & z\to-\infty\\
e^{-i\omega z}+\rho_{\omega l}^{in}\,e^{i\omega z},\,\,\,\,\, & z\to\infty
\end{cases}\nonumber \\
r\,\sqrt{4\pi\omega}\,\bar{\psi}_{\omega l}^{up}(z)\equiv\psi_{\omega l}^{up}\left(z\right)=\begin{cases}
e^{i\omega z}+\rho_{\omega l}^{up}\,e^{-i\omega z},\,\,\,\,\,\, & z\to-\infty\\
\tau_{\omega l}^{up}\,e^{i\omega z}, & z\to\infty
\end{cases}\label{eq: Etrrnal static - Basic solutions bounadry conditions}
\end{gather}
where $\tau_{\omega l}$, $\rho_{\omega l}$, represent the transmission
and reflection amplitudes. 

The \emph{Boulware state} is the vacuum state that is naturally associated
to the $\psi_{\omega l}^{in},\psi_{\omega l}^{up}$ modes. Thus, applying
the angular splitting in this state is almost identical to the prescription
given above. The only thing that requires modification is Eq. (\ref{eq: E omega l defined}),
which should now contain the two sets of modes:
\begin{equation}
E_{\omega l}\left(z\right)\equiv\left|\bar{\psi}_{\omega l}^{in}\left(z\right)\right|^{2}+\left|\bar{\psi}_{\omega l}^{up}\left(z\right)\right|^{2}.\label{eq: Etrrnal static - E(omega,l,r)}
\end{equation}

We point out that in the eternal case too the leading order large-$\omega$
contribution of the modes is independent of $l$. As can be seen in
Appendix \ref{sec: Large w approximation} (see end of Sec. \ref{subsec:Static-eternal-BH}),
both $\left|\bar{\psi}_{\omega l}^{in}\right|^{2}$ and $\left|\bar{\psi}_{\omega l}^{up}\right|^{2}$
are dominated by $1/(4\pi r^{2}\omega)$; In fact their sum 
\begin{equation}
E_{\omega l}\cong\frac{1}{2\pi r^{2}\omega}\label{eq: large-w Ewl}
\end{equation}
is the same as in the regular-center case, Eq. (\ref{eq: Omega integral - psi2 leading order}).
Hence the method of regularizing the $\omega$-integral by $l=0$
subtraction works equally well in the eternal case. Thus we again
define $F\left(l,z\right)$ according to Eq. (\ref{eq: Omega integral - Define Fl}),
and then proceed with the calculation of $\left\langle \phi^{2}\right\rangle _{ren}$
just as described in the previous subsection. 

To calculate $\left\langle \phi^{2}\right\rangle _{ren}$ in the Unruh
or Hartle-Hawking state, in the right-hand side of Eq. (\ref{eq: Etrrnal static - E(omega,l,r)})
one simply multiplies $\left|\bar{\psi}_{\omega l}^{up}\right|^{2}$,
and in Hartle-Hawking state also $\left|\bar{\psi}_{\omega l}^{in}\right|^{2}$,
by the factor $\coth(\pi\omega/\kappa)$, where $\kappa$ is the BH's
surface gravity, as described in Eqs. (\ref{eq: Ustate},\ref{eq: HHstate})
below.

\subsection{The time-dependent spherically symmetric case\label{subsec: The-general-spherically-symmetric-case}}

Generalizing the method presented in Sec. \ref{subsec: The-stationary-case}
we now consider the generic, dynamical, asymptotically-flat spherically-symmetric
metric (\ref{eq: Basic PS - Metric}), along with the field decomposition
given in Eq. (\ref{classical field decompose}). We start with the
case of a spacetime with a regular center, treating the eternal-BH
case later on. The static mode functions $\bar{\psi}_{\omega l}(z)$
are now replaced by $\bar{\Psi}_{\omega l}(t,z)e^{i\omega t}$. Therefore
the time-dependent analog of the point-splitting expression for $\left\langle \phi^{2}\left(x\right)\right\rangle _{ren}$
(summed over $m$), Eq. (\ref{eq: Omega integral - phi2 two limits}),
is
\begin{equation}
\left\langle \phi^{2}\right\rangle _{ren}=\lim_{\varepsilon\to0}\left[\lim_{\delta\to0}\hbar\sum_{l=0}^{\infty}\frac{2l+1}{4\pi}P_{l}\left(\cos\varepsilon\right)\int_{0}^{\infty}\bar{\Psi}_{\omega l}\left(t,z\right)\bar{\Psi}_{\omega l}^{*}\left(t+\delta,z\right)d\omega-G_{DS}\left(\varepsilon\right)\right].\label{eq: General SS - phi2 two limits}
\end{equation}
The biscalar $\sigma\left(\varepsilon\right)$ and the counter-term
$G_{DS}\left(\varepsilon\right)$ take the same form as in Eqs. (\ref{eq: Omega integral - sigma})-(\ref{eq: Omega integral - Gds(epsilon)}),
except that the parameters are now functions of both $t$ and $z$.
In particular $\tilde{c}$ and $c,d$ are now given by 
\[
\tilde{c}\left(t,z\right)=\frac{r^{2}}{24\Gamma}\left(\dot{r}^{2}-r'^{2}\right)\,\,,
\]
\begin{align*}
c\left(t,z\right) & =\frac{\left(1/6-\xi\right)}{8\pi^{2}r^{2}\Gamma^{3}}\left\{ -2\Gamma^{3}+r^{2}\left(\dot{\Gamma}^{2}-\Gamma'^{2}\right)+r^{2}\Gamma\left(-\ddot{\Gamma}+\Gamma''\right)\right.\\
 & +\left.2\Gamma^{2}\left(-\dot{r}^{2}+r'^{2}\right)+4r\Gamma^{2}\left(-\ddot{r}+r''\right)\right\} \,\,\,,
\end{align*}
\[
d\left(t,z\right)=\frac{1}{48\pi^{2}r\Gamma}\left(\ddot{r}-r''\right)\,\,.
\]
(The parameter $a$ is unchanged.) 

The large-$\omega$ asymptotic behavior of $|\bar{\Psi}_{\omega l}|$
is addressed in Appendix \ref{sec: Large w approximation}. In the
case of a background with regular center (either static or time dependent),
the leading-order term is given in Eq. (\ref{eq:non-eternal summary}).
Its integral over $\omega$ diverges logarithmically, but again it
is independent of $l$. Hence, here too we subtract and add the $l=0$
mode, after which the integral over $\omega$ in Eq. (\ref{eq: General SS - phi2 two limits})
takes the form 
\begin{multline*}
\int_{0}^{\infty}\left[\bar{\Psi}_{\omega l}\left(t,z\right)\bar{\Psi}_{\omega l}^{*}\left(t+\delta,z\right)-\bar{\Psi}_{\omega,l=0}\left(t,z\right)\bar{\Psi}_{\omega,l=0}^{*}\left(t+\delta,z\right)\right]d\omega\\
+\int_{0}^{\infty}\bar{\Psi}_{\omega,l=0}\left(t,z\right)\bar{\Psi}_{\omega,l=0}^{*}\left(t+\delta,z\right)d\omega.
\end{multline*}
The second integral is independent of $l$, hence as explained above
it yields zero contribution upon summation over $l$ (for finite $\varepsilon$).
The first integral converges even for $\delta=0$, so we simply insert
the limit $\delta\to0$ in the integrand. Thus, the integral over
$\omega$ in the above expression for $\left\langle \phi^{2}\right\rangle _{ren}$
reduces to 
\[
F\left(l,t,z\right)\equiv\int_{0}^{\infty}d\omega\left[E_{\omega l}\left(t,z\right)-E_{\omega,l=0}\left(t,z\right)\right],
\]
where $E_{\omega l}\left(t,z\right)$ is defined as the integrand
at the coincide: 
\begin{equation}
E_{\omega l}\left(t,z\right)\equiv\left|\bar{\Psi}_{\omega l}\left(t,z\right)\right|^{2},\label{eq: General SS - E(omega,l,r,t)}
\end{equation}
similar to Eq. (\ref{eq: E omega l defined}). Equation (\ref{eq: General SS - phi2 two limits})
now takes the form 
\[
\left\langle \phi^{2}\left(x\right)\right\rangle _{ren}=\lim_{\varepsilon\to0}\left[\hbar\sum_{l=0}^{\infty}\frac{2l+1}{4\pi}P_{l}\left(cos\,\varepsilon\right)F\left(l,t,z\right)-G_{DS}\left(\varepsilon\right)\right],
\]
similar to Eq. (\ref{eq: Omega integral - ph2 final res}) in the
static case.

From here on the method exactly follows the line described in Sec.
\ref{subsec: The-stationary-case}, except that all the $z$-dependent
quantities will now depend on $t$ as well. Summarizing the main expressions
\[
\left\langle \phi^{2}\left(x\right)\right\rangle _{ren}=\hbar\lim_{l\to\infty}H\left(l,t,z\right)+W\left(t,z\right),
\]
where
\[
H\left(l,t,z\right)\equiv\sum_{k=0}^{l}\frac{2k+1}{4\pi}\left[F_{reg}\left(k,t,z\right)-F_{reg}\left(l,t,z\right)\right],
\]
\[
W\left(t,z\right)\equiv-\hbar\left[\left(\ln\left(2\mu r\right)+\gamma\right)c\left(t,z\right)+d\left(t,z\right)\right],
\]
and 
\[
F_{reg}\left(l,t,z\right)\equiv F\left(l,t,z\right)-F_{sing}\left(l,t,z\right),
\]
\[
F_{sing}\left(l,t,z\right)=-8\pi a\left(t,z\right)h\left(l\right)+2\pi c\left(t,z\right)\Lambda\left(l\right).
\]

\subsubsection*{The dynamical eternal-BH case\label{subsec: Two-ended dynamic}}

Similar to the static case, in a dynamical eternal-BH background there
are two sets of basis solutions $\bar{\Psi}_{\omega l}^{in}\left(t,z\right),\bar{\Psi}_{\omega l}^{up}\left(t,z\right)$.
The initial conditions for these solutions are easy to express in
double-null coordinates: 
\begin{equation}
\lim_{pni}\Psi_{\omega l}^{in}\left(u,v\right)=e^{-i\omega v},\,\,\,\lim_{ph}\Psi_{\omega l}^{in}\left(u,v\right)=0\label{eq: Etrrnal dynamic - Boundary conditions-in}
\end{equation}
and
\begin{equation}
\lim_{ph}\Psi_{\omega l}^{up}\left(u,v\right)=e^{-i\omega u},\,\,\,\lim_{pni}\Psi_{\omega l}^{up}\left(u,v\right)=0\,,\label{eq: Etrrnal dynamic - Boundary conditions-up}
\end{equation}
where ``pni'' and ``ph'' stand for ``past null infinity'' and
``past horizon'' respectively. \footnote{The vacuum state naturally associated to this set of modes is the
(time-dependent analog of the) Boulware state. } Thus the only change required from the non-eternal dynamical case
described above is in Eq. (\ref{eq: General SS - E(omega,l,r,t)}),
which is now replaced by 
\[
E_{\omega l}\left(t,z\right)\equiv\left|\bar{\Psi}_{\omega l}^{in}\left(t,z\right)\right|^{2}+\left|\bar{\Psi}_{\omega l}^{up}\left(t,z\right)\right|^{2}.
\]

Note that in a static background the Eddington-like null coordinates
$u,v$ are naturally defined by the time translation symmetry, but
in a dynamical eternal-BH background this definition no longer holds.
The coordinate $v$ is still naturally defined by asymptotic flatness
at PNI (and indeed we assume this definition of $v$ throughout),
but $u$ is no longer uniquely defined. The boundary conditions in
Eq. (\ref{eq: Etrrnal dynamic - Boundary conditions-up}) thus induce
a vacuum state that depends on the specific choice of the $u$ coordinate,
via the definition of the $\Psi_{\omega l}^{up}$ modes. This ambiguity
in the choice of vacuum does not arise in the non-eternal case (and
certainly not in the static case). \footnote{One may choose to define $u$ via asymptotic flatness at FNI, but
it is not clear if this will always be the most convenient choice.
Another natural candidate is the log of the affine parameter along
the past horizon. In the non-eternal case a natural choice of $u$
may arise from the requirement that the center of symmetry would be
placed at $z=0$. }

\section{Application to the Schwarzschild case\label{sec: Schwarzschild}}

We now demonstrate the implementation of the angular-splitting variant
by calculating $\left\langle \phi^{2}\left(x\right)\right\rangle _{ren}$
in the exterior of a Schwarzschild spacetime, for a massless scalar
field. \footnote{Note that in Schwarzschild $\left\langle \phi^{2}\right\rangle _{ren}$
does not depend on $\xi$ because $R=0$.} We do this first in the Boulware vacuum, and later on in Sec. \ref{subsec: Unruh and Hartle-Hawking}
we also give results for the Unruh and Hartle-Hawking states. The
Schwarzschild metric is
\[
ds^{2}=-\left(1-\frac{2M}{r}\right)dt^{2}+\left(1-\frac{2M}{r}\right)^{-1}dr^{2}+r^{2}d\Omega^{2},
\]
where $M$ is the BH mass. Defining the usual tortoise coordinate
\[
r_{*}=r+2M\ln\left(\frac{r}{2M}-1\right)
\]
the metric can be written as
\[
ds^{2}=\left(1-\frac{2M}{r}\right)\left(-dt^{2}+dr_{*}^{2}\right)+r^{2}d\Omega^{2}.
\]
Thus, in the Schwarzschild case the coordinates $z$ used below can
be replaced by the more familiar symbol $r_{*}$. Also, since in the
exterior of Schwarzschild $r(r_{*})$ is a monotonically increasing
function, it is now possible (and it is often convenient) to use $r$
as a radial variable. (Which is not the case in the general static
case, where $r(z)$ need not be monotonic.)

The radial equation for the modes is as given in Eq. (\ref{eq: Stationary case - field eq-1}), 

\begin{equation}
\psi''_{\omega l}\left(r\right)=\left[V_{l}\left(r\right)-\omega^{2}\right]\psi_{\omega l}\left(r\right),\label{eq: Sch - radial equation}
\end{equation}
where a prime denotes $d/dz\equiv d/dr_{*}$, and the effective potential
in the Schwarzschild case takes the form

\[
V_{l}\left(r\right)=\left(1-\frac{2M}{r}\right)\left[\frac{l\left(l+1\right)}{r^{2}}+\frac{2M}{r^{3}}\right].
\]

The Schwarzschild geometry describes an eternal BH, so there are two
sets of basis solutions $\bar{\psi}_{\omega l}^{in},\bar{\psi}_{\omega l}^{up}$
and the scheme is executed following Sec. \ref{subsec: Two-ended static}.
For the Schwarzschild background (and for $m=0$) the parameters $a\left(r\right),c\left(r\right),d\left(r\right)$
take the simple forms 
\[
a\left(r\right)=\frac{1}{16\pi^{2}r^{2}}\,,\,\,c\left(r\right)=0,\,\,d\left(r\right)=-\frac{M}{24\pi^{2}r^{3}}\,\,.
\]
For reference we also give here the explicit expressions for $F_{sing}\left(l,r\right)$
and $W\left(r\right)$:
\begin{equation}
F_{sing}\left(l,r\right)=-\frac{1}{2\pi r^{2}}\,h\left(l\right)\,\,,\,\,\,\,\,W\left(r\right)=\hbar\,\frac{M}{24\pi^{2}r^{3}}\,.\label{eq: Sch - Fsing =000026 W}
\end{equation}
Then $\left\langle \phi^{2}\right\rangle _{ren}$ is given by Eq.
(\ref{eq: Blind-spots - phi2 final}).

\subsection{Numerical implementation in Schwarzschild\label{subsec:Numerical-implementation-in}}

The radial equation (\ref{eq: Sch - radial equation}) was numerically
solved for $\psi_{\omega l}\left(r\right)$ using MATHEMATICA's ODE
solver. It was solved for 21 different $l$ values ($0\leq l\leq20$),
and for each $l$ in the range $\omega\in\left[0,10\right]$, with
a uniform spacing $d\omega=1/1000$, \footnote{Note that in most cases one does not need to go up to $l=20$, usually
the convergence is much faster. We found, however, that for large
$r$ values in the Hartle-Hawking state the convergence in $l$ is
slower so that we needed the modes up to $l=20$ to keep the high
accuracy. We also note that a spacing of $d\omega=1/1000$ is really
unnecessary and one can get excellent accuracy even with $d\omega=1/100$.} namely $\sim2\cdot10^{5}$ modes. In all graphs below we use units
where $M=1$, in addition to $G=c=1$. 

\begin{figure}
\subfloat[Solid curve: the numerically calculated $E_{\omega,l=1}$, as defined
in Eq. (\ref{eq: Etrrnal static - E(omega,l,r)}), evaluated at $r=6M$.
Its large-$\omega$ asymptotic behavior $\propto\omega^{-1}$ is evident
from the dashed line which is $\omega E_{\omega,l=1}$. \label{fig: figure 1a}]{\begin{centering}
\includegraphics[bb=20bp 180bp 560bp 620bp,clip,scale=0.4]{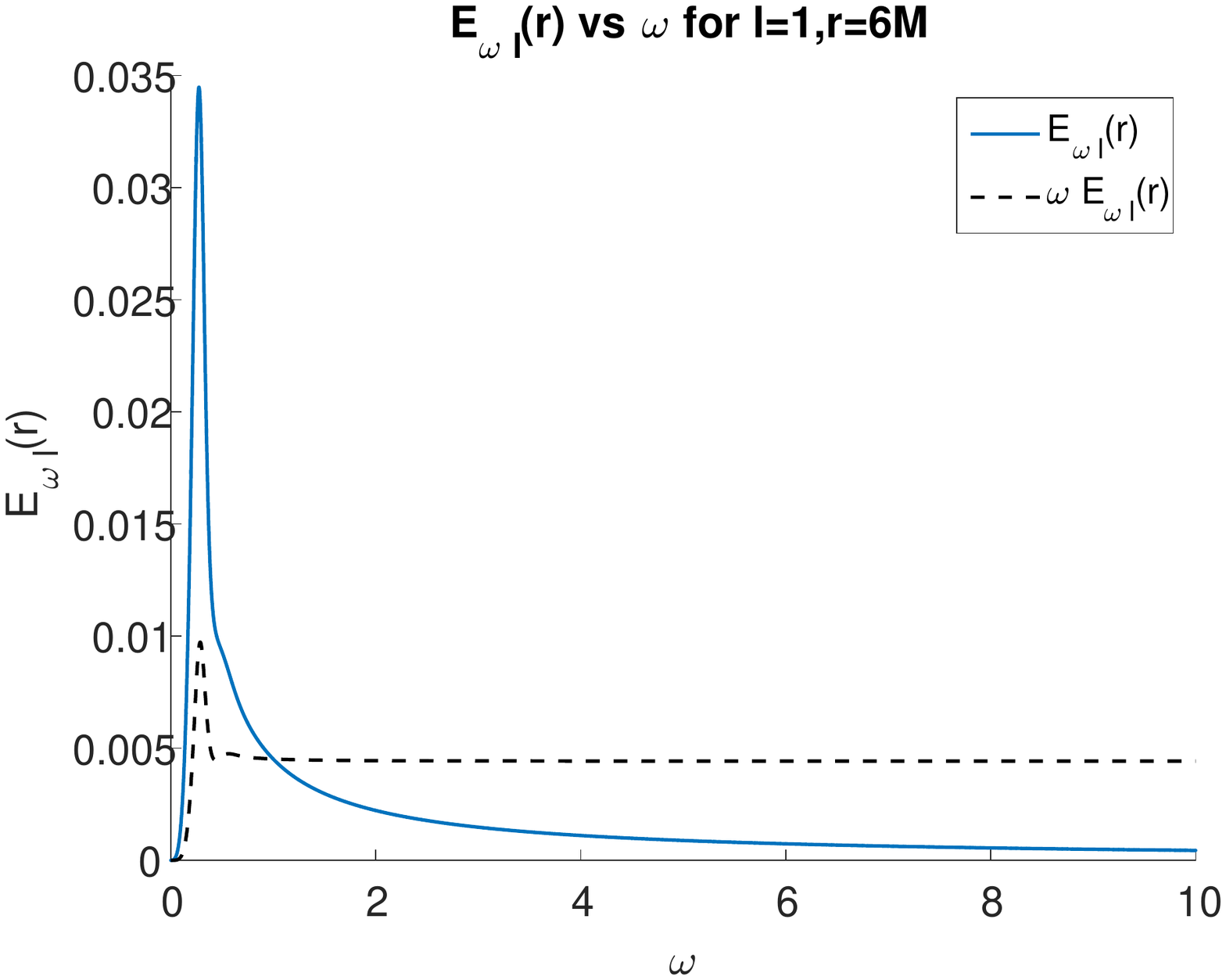}
\par\end{centering}
}\enskip{}\hfill{}\enskip{}\subfloat[Solid curve: the subtraction $E_{\omega,l=1}-E_{\omega,l=0}$ at $r=6M$.
It decays as $\omega^{-3}$, as seen from the dashed line which is
the same subtraction multiplied by $100\omega^{3}$.\label{fig: figure 1b}]{\begin{centering}
\includegraphics[bb=20bp 180bp 560bp 620bp,clip,scale=0.4]{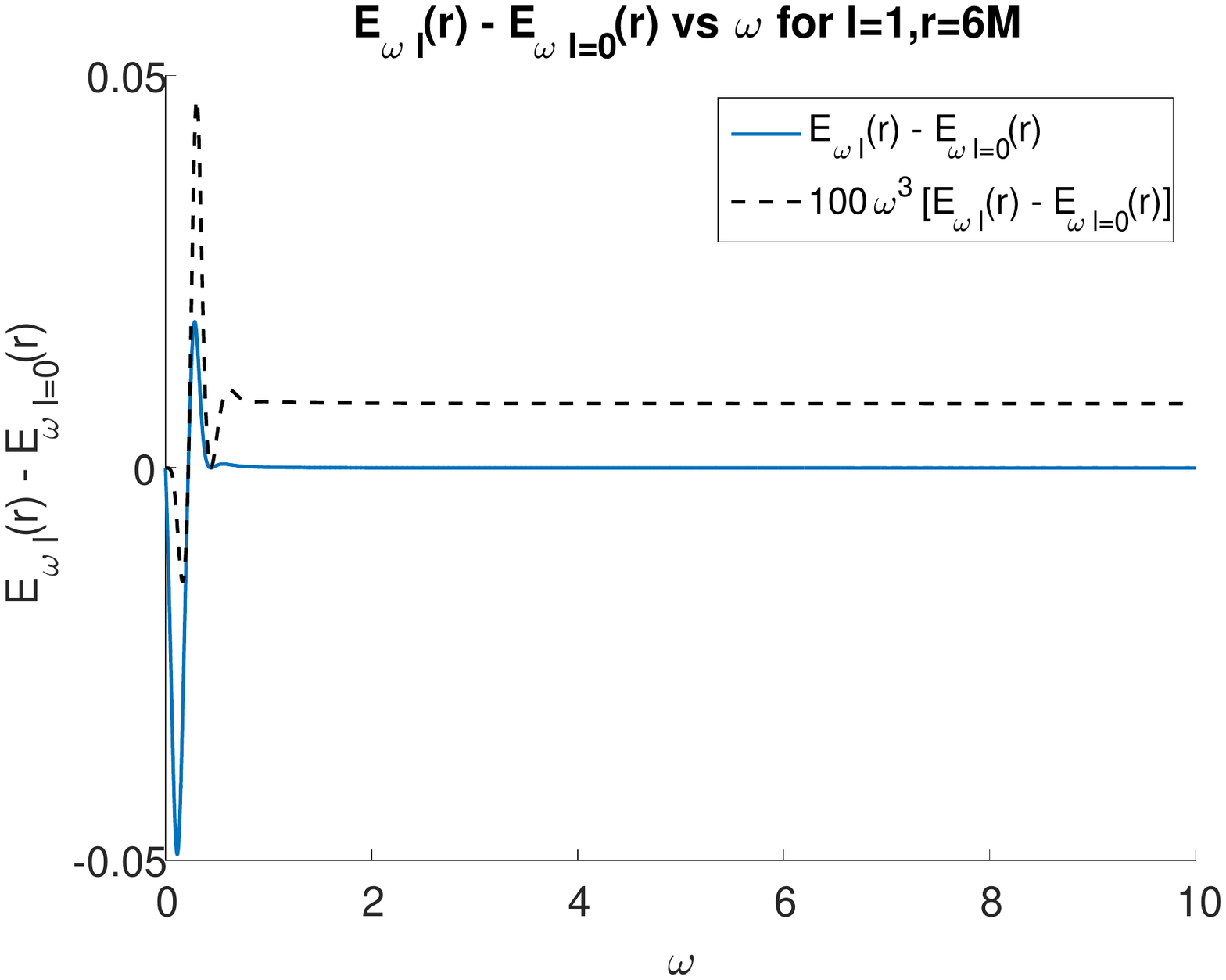}
\par\end{centering}
}

\caption{}
\end{figure}

As an example we follow the calculation for $r=6M$ in Boulware state,
illustrating the various stages of the regularization process. Figure
\ref{fig: figure 1a} displays $E_{\omega l}$ for $l=1$. This quantity
behaves like $1/\omega$ at large $\omega$ (see dashed curve), so
its integral would diverge at infinity. It is regularized according
to Eq. (\ref{eq: Omega integral - Define Fl}) by subtracting from
it the $l=0$ mode; the resultant integrand behaves as $1/\omega^{3}$
for large $\omega$, as seen in Fig. \ref{fig: figure 1b} (dashed
curve).

Although the $\omega$-integral of $E_{\omega l}\left(r\right)-E_{\omega,l=0}\left(r\right)$
is convergent, it does not converge sufficiently fast, due to the
$\omega^{-3}$ tail. This would bring up the need for a much longer
range in $\omega$ (and hence a much larger number of modes) in order
to achieve a sufficient accuracy of the integral. To overcome this
difficulty, we carried the large-$\omega$ expansion of $\psi_{\omega l}\left(r\right)$
up to order $\omega^{-8}$, and used this analytical approximation
for evaluating the integral from $\omega=10$ to infinity. This expansion
is described in Appendix \ref{sec: Large w approximation}, see in
particular Eqs. (\ref{eq:Eternal}-\ref{eq:coefficients}). This is
one way to obtain a more accurate result with a limited range in $\omega$,
but one can also think of other ways.

\begin{figure}
\subfloat[The solid curve represents $F\left(l,r=6M\right)$, calculated according
to Eq. (\ref{eq: Omega integral - Define Fl}). The dashed curve is
the analytic function $F_{sing}\left(l,r=6M\right)$, given in Eq.
(\ref{eq: Sch - Fsing =000026 W}). \label{fig: figure 2a}]{\begin{centering}
\includegraphics[bb=20bp 180bp 560bp 620bp,clip,scale=0.4]{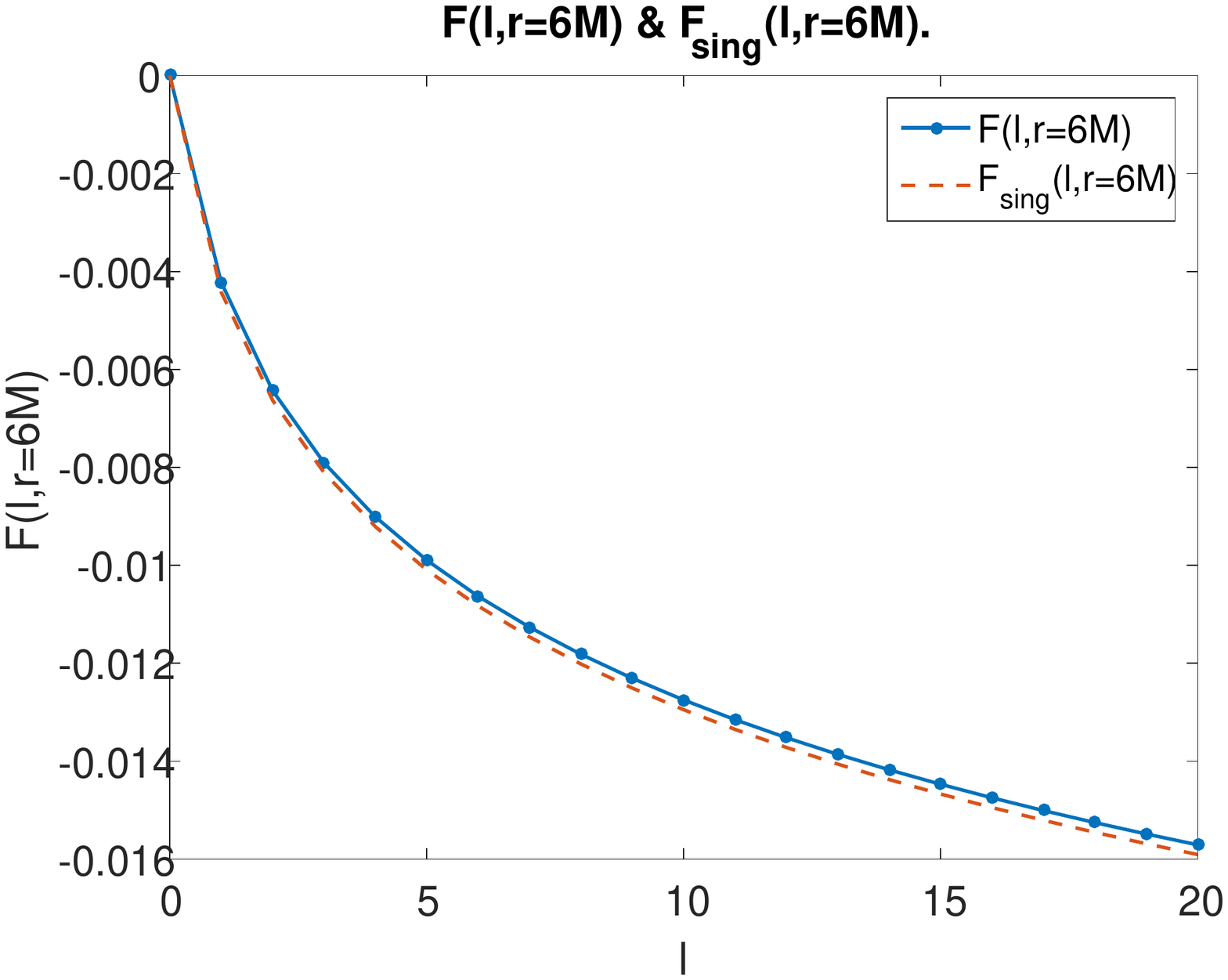}
\par\end{centering}
}\enskip{}\hfill{}\enskip{}\subfloat[The graph displays $F_{reg}\left(l,r=6M\right)$, which is simply
the difference between the two curves in Fig. \ref{fig: figure 2a}.
It rapidly approaches a constant, the ``blind-spot'' discussed in
Sec. \ref{subsec: Blind-spots}.\label{fig: figure 2b}]{\begin{centering}
\includegraphics[bb=20bp 180bp 560bp 620bp,clip,scale=0.4]{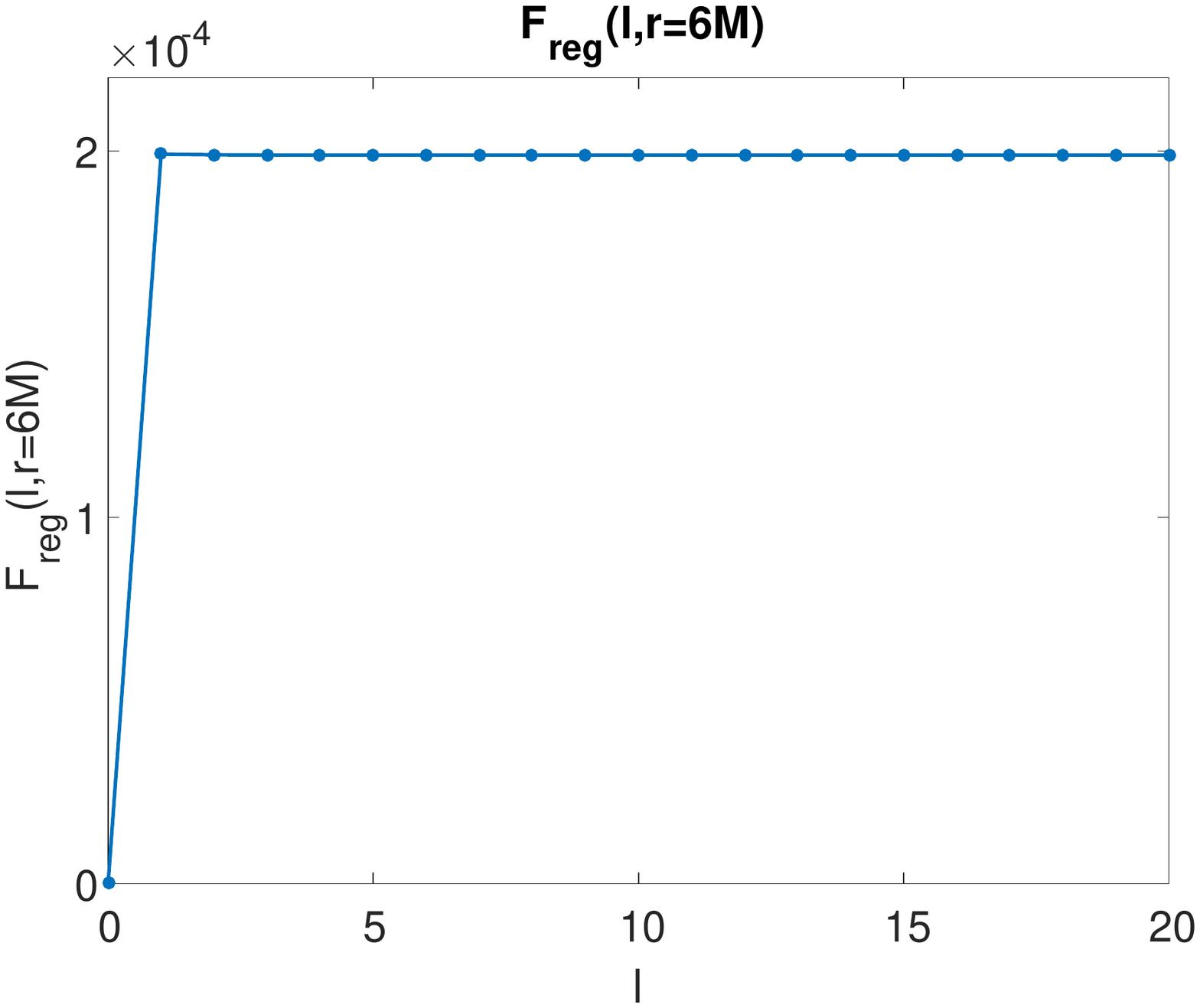}
\par\end{centering}
}

\caption{}
\end{figure}

The result of the integration over $\omega$ is $F\left(l,r\right)$,
seen in Fig. \ref{fig: figure 2a}. As expected it behaves logarithmically
for large $l$ (like $F_{sing}\left(l,r\right)$), and the sum over
$l$ of $\left(2l+1\right)F\left(l,r\right)$ is of course divergent.
The subtraction of $F_{sing}\left(l,r\right)$ according to Eq. (\ref{eq: l sum - Freg})
eliminates most of the divergent part of $F\left(l,r\right)$. The
result is $F_{reg}\left(l,r\right)$, which asymptotically behaves
like a constant (Fig. \ref{fig: figure 2b}). This fits our understanding
of the blind-spot and matches the structure of Eq. (\ref{eq: Blind-spots - Freg}).

\begin{figure}
\subfloat[The sequence $H\left(l,r\right)$ constructed according to Eq. \label{fig: Figure 3a}
(\ref{eq: Blind-spots - H}). The fast convergence is evident.]{\begin{centering}
\includegraphics[bb=20bp 180bp 560bp 620bp,clip,scale=0.4]{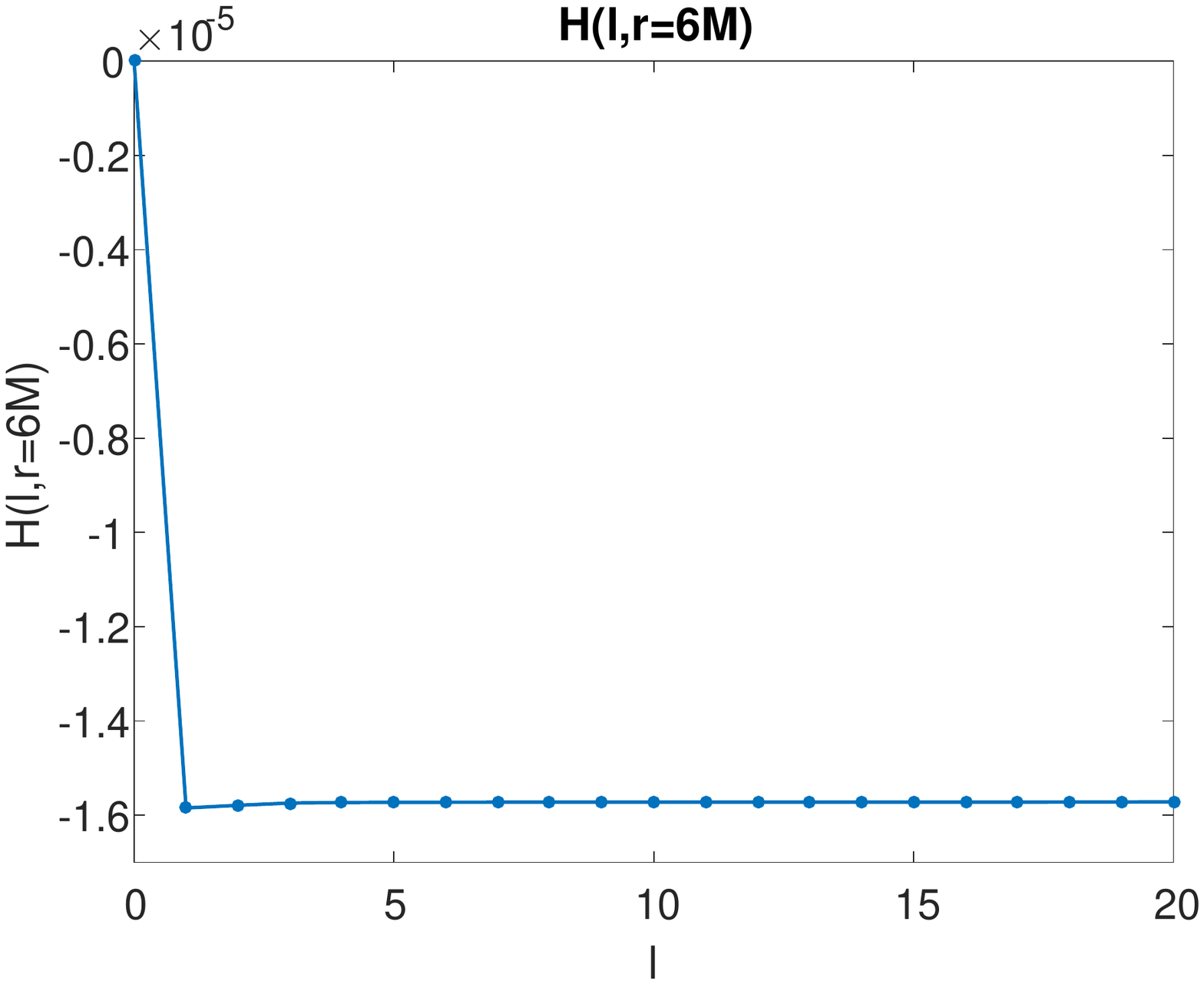}
\par\end{centering}
}\enskip{}\hfill{}\enskip{}\subfloat[A close-up on the plateau in Fig. \ref{fig: Figure 3a}. In this zoom
one can also notice the growth of numerical error at large $l$ (say
$l>16$). The red cross indicates the estimated optimal $l$ value
(for numerically evaluating the large-$l$ limit of $H\left(l,r\right)$;
$l=12$ in the present case). This estimated optimal $l$ is automatically
selected by an algorithm that assesses where the numerical error starts
to grow.\label{fig: Figure 3b}]{\begin{centering}
\includegraphics[bb=20bp 180bp 560bp 620bp,clip,scale=0.4]{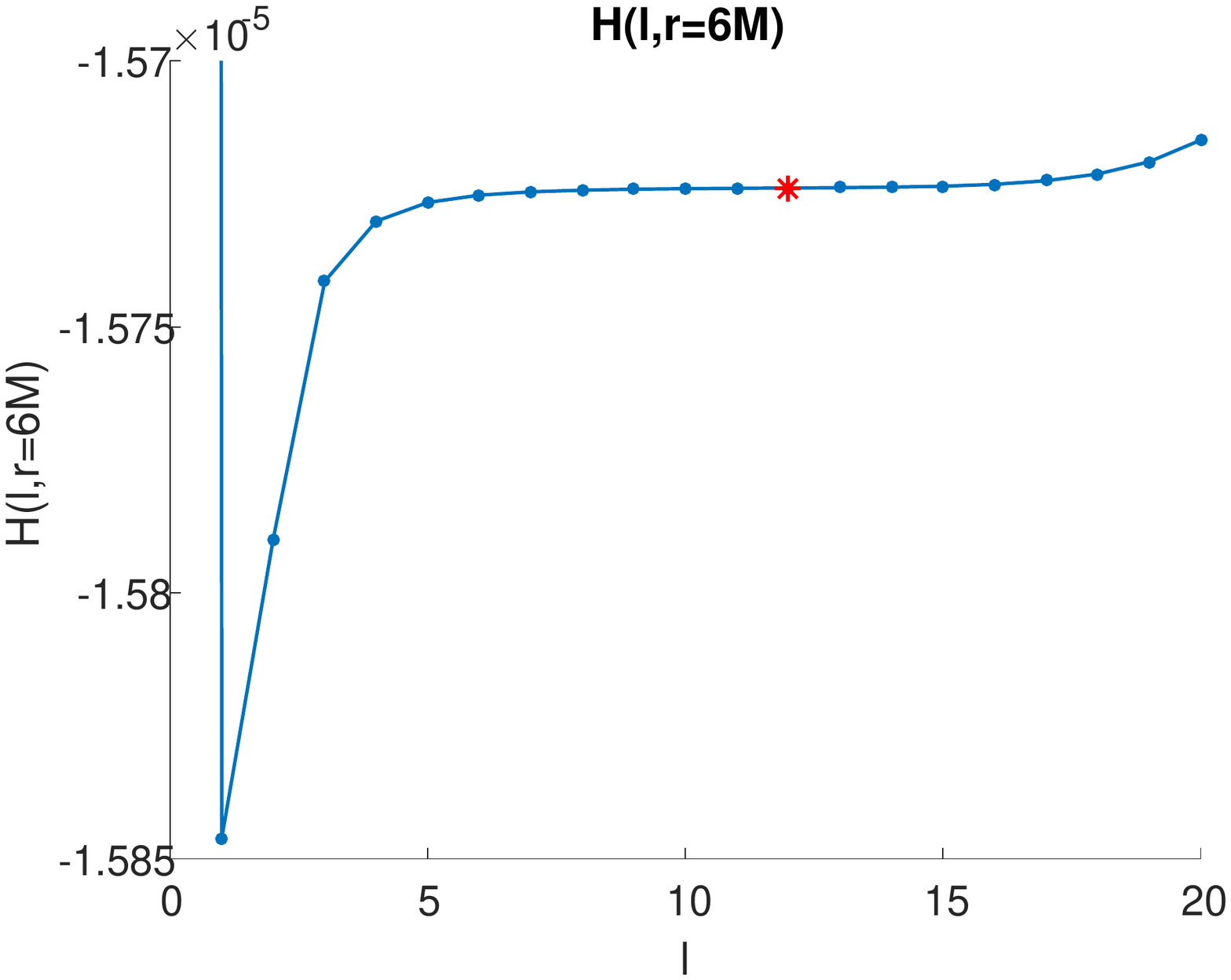}
\par\end{centering}
}

\caption{}
\end{figure}

Next we self-cancel the blind spot by constructing $H\left(l,r\right)$
according to Eq. (\ref{eq: Blind-spots - H}). It is clear from Fig.
\ref{fig: Figure 3a} that the sequence $H\left(l,r\right)$ indeed
converges as anticipated; furthermore, it converges very fast as can
be seen from the zoom in Fig. \ref{fig: Figure 3b}. After the large-$l$
limit of $H\left(l,r\right)$ is computed, $\left\langle \phi^{2}\right\rangle _{ren}$
is calculated according to Eq. (\ref{eq: Blind-spots - phi2 final}). 

\begin{figure}
\subfloat[The solid line represents the results for $\left\langle \phi^{2}\right\rangle _{ren}$
in the Boulware state, calculated using the new angular-splitting
variant. There is excellent agreement with both the results obtained
using the $t$-splitting variant (the crosses) and previous results
by Anderson (the asterisks).\label{fig: Figure 4a}]{\begin{centering}
\includegraphics[bb=20bp 180bp 560bp 620bp,clip,scale=0.4]{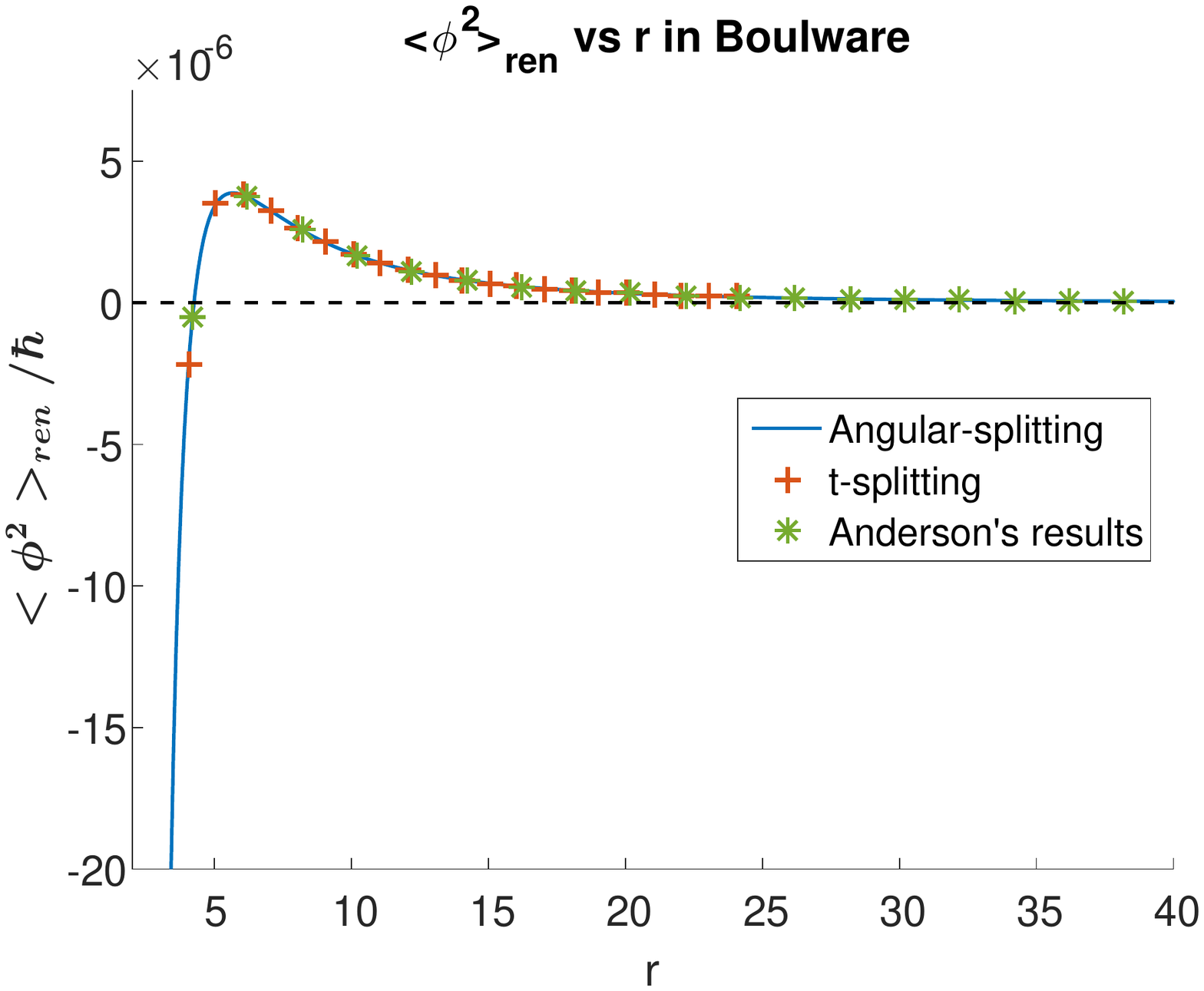}
\par\end{centering}
}\enskip{}\hfill{}\enskip{}\subfloat[A near horizon close-up of $\left(1-2M/r\right)\left\langle \phi^{2}\right\rangle _{ren}$,
 showing that it is clearly finite. Here we also give the analytic
horizon value calculated by Candelas (the dotted line). \label{fig: Figure 4b}]{\begin{centering}
\includegraphics[bb=20bp 180bp 560bp 620bp,clip,scale=0.4]{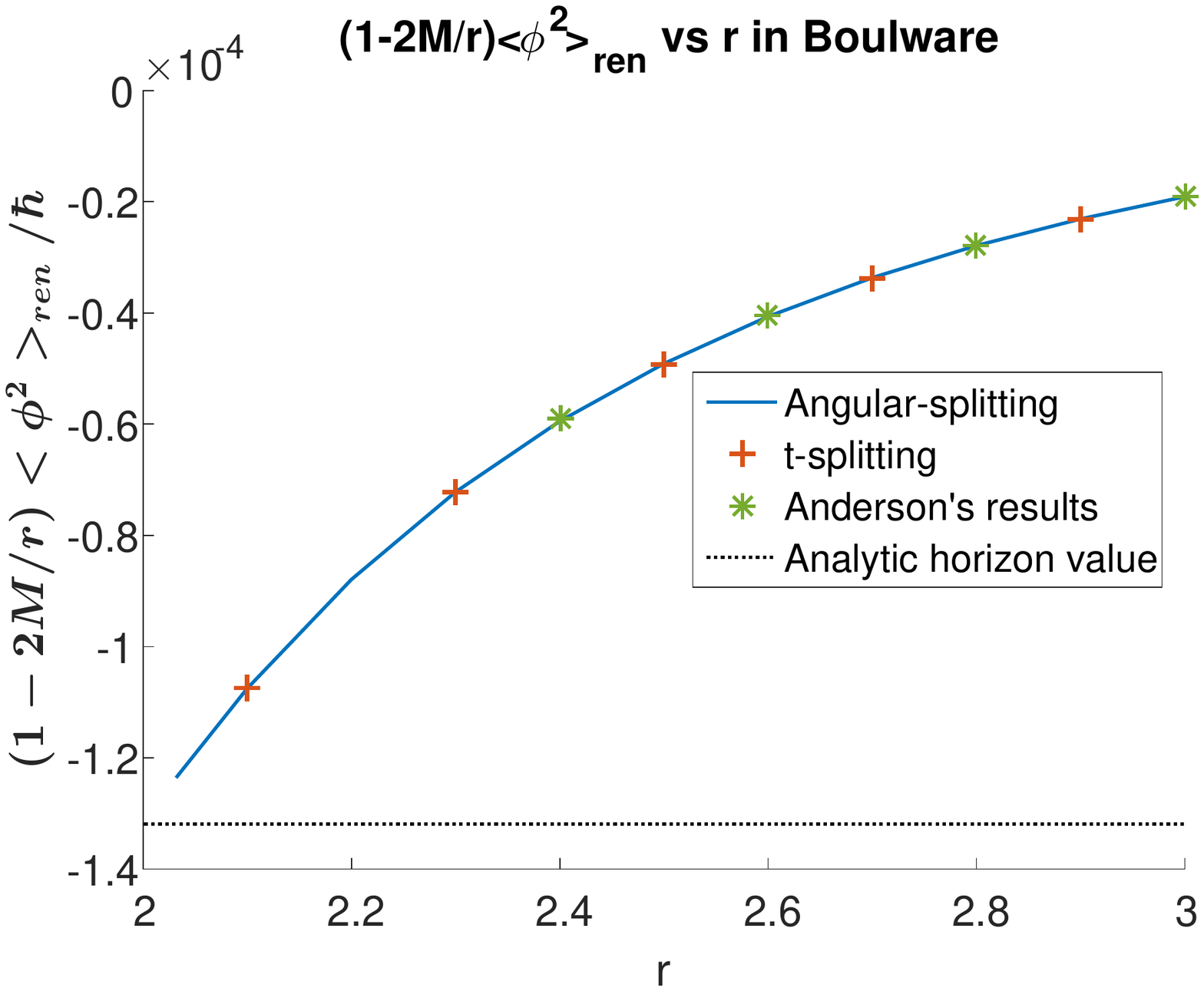}
\par\end{centering}
}

\caption{}
\end{figure}

In Fig. \ref{fig: Figure 4a} we present $\left\langle \phi^{2}\right\rangle _{ren}$
as a function of $r$ in Boulware state, for various $r$ values between
$2.03M$ and $40M$ \footnote{Our numerical results get close to the horizon up to $r=2.001M$.
However, in the Boulware state we give the results only up to $2.03M$
because closer than that the accuracy deteriorates rapidly. This is
expected due to the divergence of the Boulware state on approaching
the horizon.}, and compare it to previous results calculated using the $t$-splitting
variant \cite{Levi =000026 Ori - 2015 - t splitting regularization}.
The two variants admit excellent agreement and the difference is typically
of order one part in $10^{4}$. \footnote{This does not necessarily mean that the error in the numerically evaluated
$\left\langle \phi^{2}\right\rangle _{ren}$ is that small. The error
may turn out to be larger than the difference between the two variants.
For example, a numerical error in evaluating the contribution from
any mode that we include in both splittings, will yield same error
in the two variants and will not show up in their difference. Nevertheless,
from various indicators we estimate that the error is typically around
one part in $10^{3}$ or smaller. } We also compare our results to the ones calculated by Anderson \cite{Anderson (private)}
using an entirely different method (WKB expansion in the euclidean
sector), which also admits very good agreement. The near-horizon asymptotic
behavior of $\left\langle \phi^{2}\right\rangle _{ren}$ in Boulware
state is shown in Fig. \ref{fig: Figure 4b}; as expected it diverges
like $(1-2M/r)^{-1}$ so we plot $(1-2M/r)\left\langle \phi^{2}\right\rangle _{ren}$
which is finite. There is good visual agreement with the known limiting
value $-\hbar/(768\pi^{2}M^{2})$ (the dotted horizontal line in
Fig. \ref{fig: Figure 4b}) at $r=2M$, calculated analytically by
Candelas \cite{Candelas - 1979 - phi2 Schwarzschild}. 

\subsection{Unruh and Hartle-Hawking states\label{subsec: Unruh and Hartle-Hawking}}

In addition to the Boulware state, we also used the angular splitting
method to calculate $\left\langle \phi^{2}\right\rangle _{ren}$ in
the Unruh and Hartle-Hawking vacuum states. The regularization process
is the same as the one described above for the Boulware state; the
only difference is in the mode contributions. In principle each vacuum
state is established on different sets of basis modes, defined according
to different boundary conditions. Nevertheless it is not necessary
to re-solve the differential equations for the mode functions for
each state separately: One can construct the mode contributions for
the Hartle-Hawking and Unruh states from the same set of solutions
$\psi_{\omega l}^{in}\left(r\right),\psi_{\omega l}^{up}\left(r\right)$
that used us for the Boulware state. We give here the mode contributions
in the different vacuum states and refer the reader to derivation
of these relations by Christensen and Fulling \cite{Christensen =000026 Fulling - 1977}:

\begin{equation}
E_{\omega l}^{Boulware}\left(r\right)\equiv\left|\bar{\psi}_{\omega l}^{in}\left(r\right)\right|^{2}+\left|\bar{\psi}_{\omega l}^{up}\left(r\right)\right|^{2},\label{eq: Bstate}
\end{equation}
{[}which is actually the original function $E_{\omega l}$ given in
Eq. (\ref{eq: Etrrnal static - E(omega,l,r)}){]}, and 
\begin{equation}
E_{\omega l}^{Unruh}\left(r\right)\equiv\left|\bar{\psi}_{\omega l}^{in}\left(r\right)\right|^{2}+\coth\left(\frac{\pi\omega}{\kappa}\right)\left|\bar{\psi}_{\omega l}^{up}\left(r\right)\right|^{2},\label{eq: Ustate}
\end{equation}
\begin{equation}
E_{\omega l}^{H-H}\left(r\right)\equiv\coth\left(\frac{\pi\omega}{\kappa}\right)\left(\left|\bar{\psi}_{\omega l}^{in}\left(r\right)\right|^{2}+\left|\bar{\psi}_{\omega l}^{up}\left(r\right)\right|^{2}\right),\label{eq: HHstate}
\end{equation}
where $\kappa$ is the surface gravity of the black hole, in Schwarzschild
$\kappa=1/(4M)$. Notice that the regularization in $\omega$ does
not require any modification since the large-$\omega$ asymptotic
behavior is the same for the three vacuum states, as $\coth\left(\pi\omega/\kappa\right)$
exponentially approaches one at large $\omega$. 

\begin{figure}
\subfloat[The solid curve displays $\left\langle \phi^{2}\right\rangle _{ren}$
in Unruh state calculated using angular splitting. The results obtained
using the $t$-splitting variant are marked by crosses. The dashed
curve is aimed to demonstrate  the $\propto(r^{-2})$  large-$r$
asymptotic behavior of $\left\langle \phi^{2}\right\rangle _{ren}$,
by multiplying the latter by $r^{2}/10$. \label{fig: Figure 5a}]{\begin{centering}
\includegraphics[bb=20bp 180bp 560bp 620bp,clip,scale=0.4]{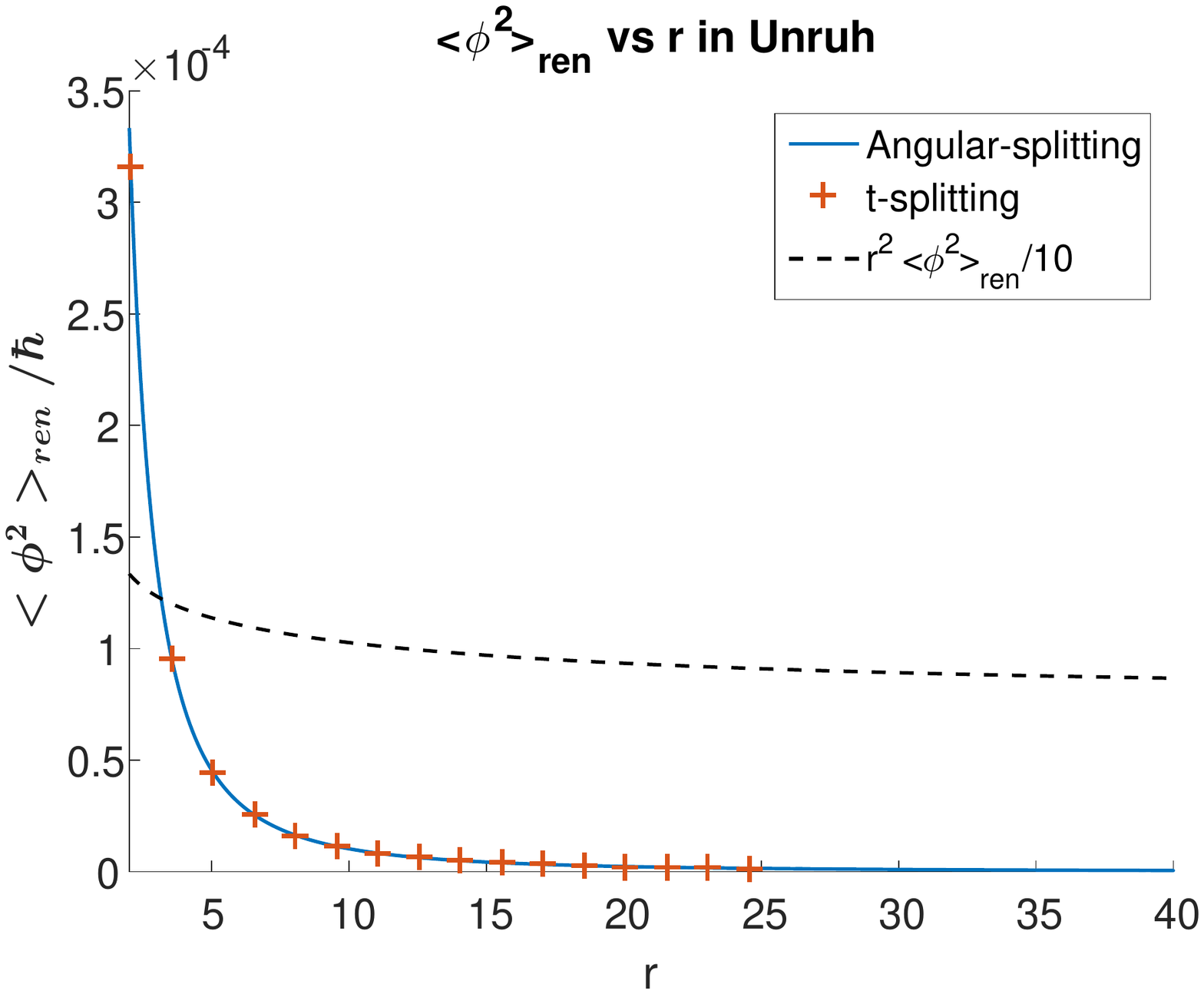}
\par\end{centering}
}\enskip{}\hfill{}\enskip{}\subfloat[A near-horizon zoom on $\left\langle \phi^{2}\right\rangle _{ren}$
in Unruh state, demonstrating its regularity at $r\to2M$.\label{fig: Figure 5b}]{\begin{centering}
\includegraphics[bb=20bp 180bp 560bp 620bp,clip,scale=0.4]{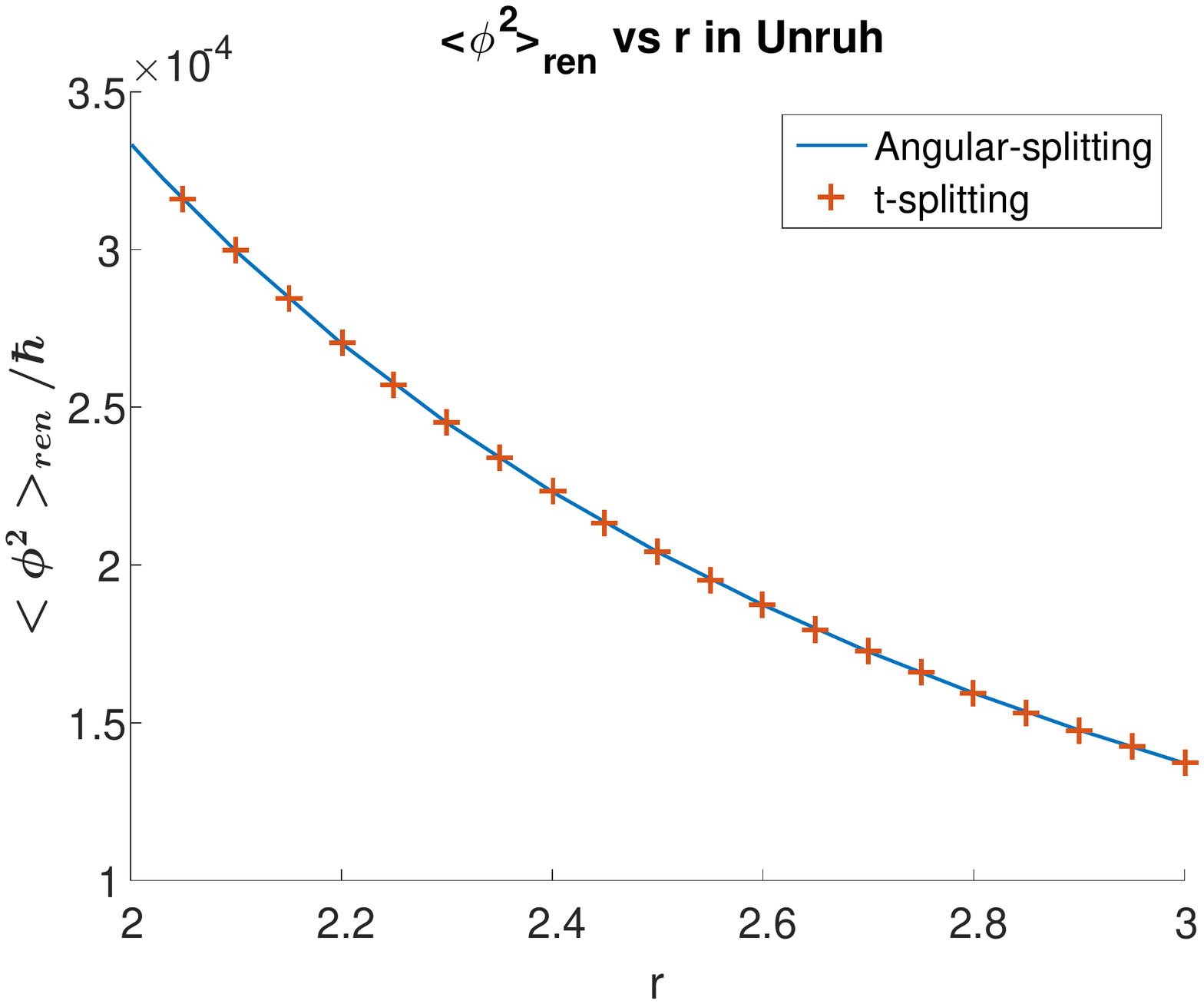}
\par\end{centering}
}

\caption{}
\end{figure}

Figure \ref{fig: Figure 5a} displays $\left\langle \phi^{2}(r)\right\rangle _{ren}$
in the Unruh state, along with results calculated using the $t$-splitting
variant. The agreement is again excellent and the difference is typically
a few parts in $10^{5}$. From this figure one might get the wrong
visual impression that $\left\langle \phi^{2}\right\rangle _{ren}$
diverges at $r\to2M$. To show that this is not the case, Fig. \ref{fig: Figure 5b}
zooms on the closer neighborhood of the horizon and demonstrates the
regularity of $\left\langle \phi^{2}(r)\right\rangle _{ren}$ as $r\to2M$. 

For future reference we also give here the horizon value extrapolated
from the points calculated near the horizon: 
\[
\left\langle \phi^{2}(r=2M)\right\rangle _{ren}^{Unruh}\approx3.336\cdot10^{-4}\frac{\hbar}{M^{2}}.
\]
In addition, the leading-order asymptotic behavior at infinity was
extrapolated and found to be 
\[
\left\langle \phi^{2}(r\to\infty)\right\rangle _{ren}^{Unruh}\approx7.763\cdot10^{-4}\frac{\hbar}{r^{2}}.
\]
Candelas gave analytic expressions for the asymptotic behavior at
the horizon and at infinity. These two asymptotic expressions depend
on the gray-body factor (see Table I in Ref \cite{Candelas - 1979 - phi2 Schwarzschild}).
Nevertheless we find that there exists a combination, which we denote
by $\chi$, that cancels the gray-body factor and yields an explicit
analytic value. This combination is 
\[
\chi=\left\langle \phi^{2}(r=2M)\right\rangle _{ren}^{Unruh}+\frac{r^{2}}{4M^{2}}\left\langle \phi^{2}(r\to\infty)\right\rangle _{ren}^{Unruh}=\frac{\hbar}{192\pi^{2}M^{2}}\,.
\]
The numerically calculated results agree with this analytic value
to about one part in $10^{4}$. 

\begin{figure}
\subfloat[The solid curve displays $\left\langle \phi^{2}\right\rangle _{ren}$
in the Hartle-Hawking state calculated using the angular-splitting
method. The results obtained from the $t$-splitting variant are marked
by crosses. The analytical asymptotic values calculated by Candelas
are represented by the two dotted horizontal lines (top line \textemdash{}
horizon; bottom line \textemdash{} infinity). \label{fig: Figure 6a}]{\begin{centering}
\includegraphics[bb=20bp 180bp 560bp 620bp,clip,scale=0.4]{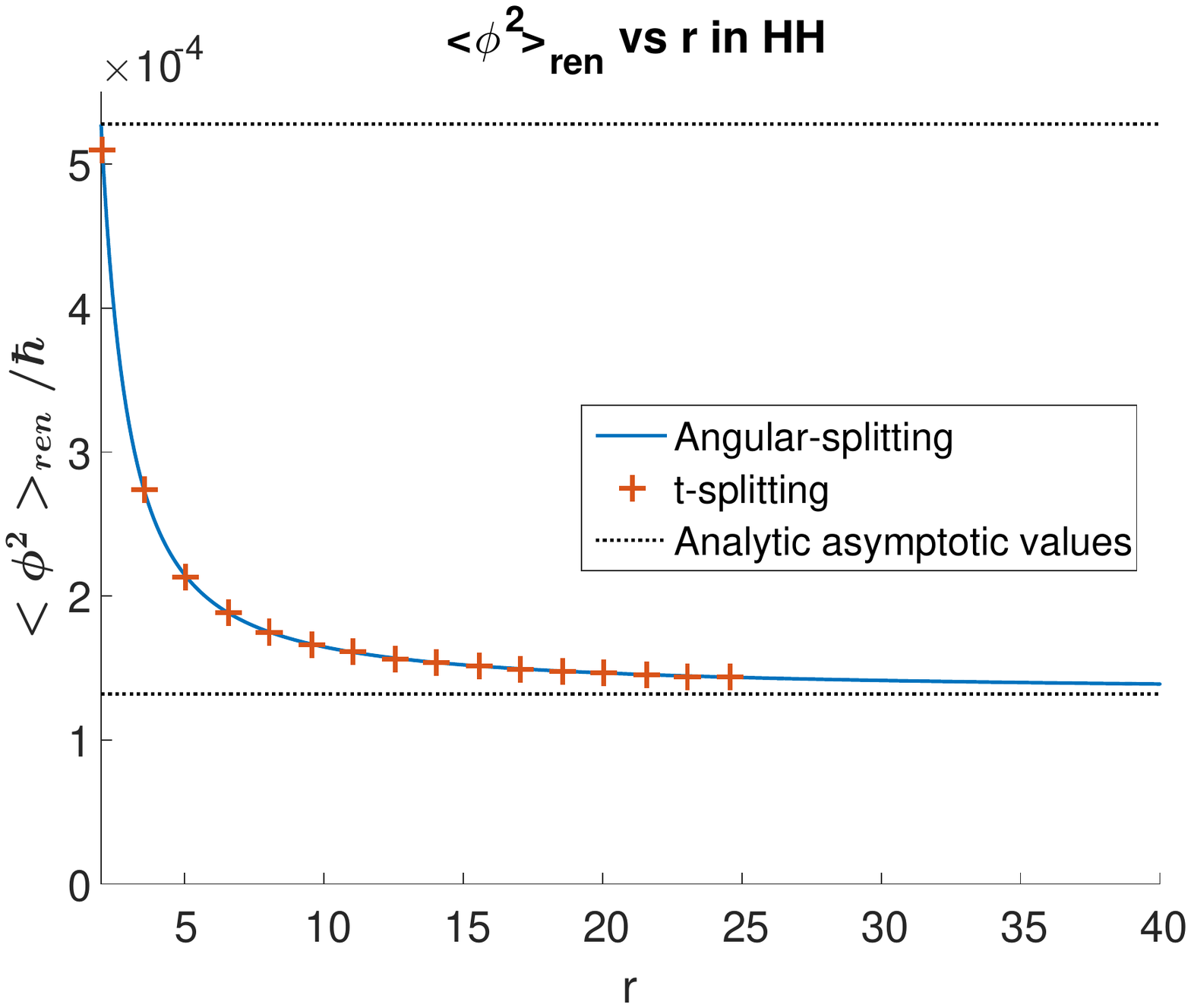}
\par\end{centering}
}\enskip{}\hfill{}\enskip{}\subfloat[A near-horizon zoom on $\left\langle \phi^{2}\right\rangle _{ren}$
in the Hartle-Hawking state. One can see the agreement with the analytical
value at the horizon calculated by Candelas (the dotted horizontal
line). \label{fig: Figure 6b}]{\begin{centering}
\includegraphics[bb=20bp 180bp 560bp 620bp,clip,scale=0.4]{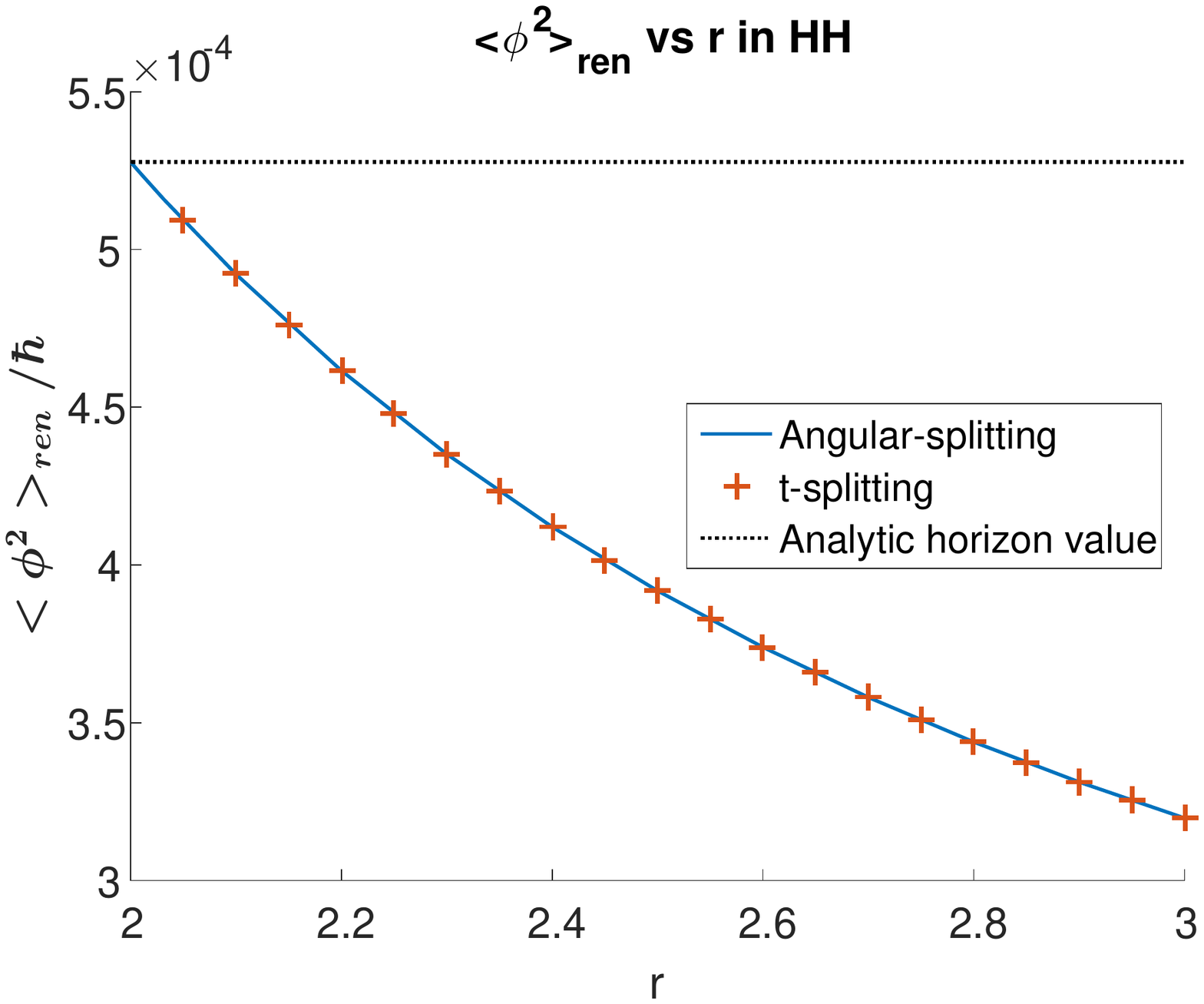}
\par\end{centering}
}

\caption{}
\end{figure}

Finally we also display $\left\langle \phi^{2}\right\rangle _{ren}$
in the Hartle-Hawking state in Fig. \ref{fig: Figure 6a}. Again the
deviations from the calculation using the $t$-splitting variant are
typically a few parts in $10^{5}$. The figure also shows the asymptotic
values analytically calculated by Candelas \cite{Candelas - 1979 - phi2 Schwarzschild}
at infinity ($\hbar/768\pi^{2}M^{2}$) and at the horizon ($\hbar/192\pi^{2}M^{2}$).
Similar to the Unruh state the value at the horizon is finite, this
is clearly visible in Fig. \ref{fig: Figure 6b}.  We have calculated
the limiting values at $r=2M$ and at $r\to\infty$, and these extrapolated
numerical values agree with Candelas' analytical results up to two
parts in $10^{5}$ and three part in $10^{4}$ respectively. 

\section{Discussion\label{sec: Discussion}}

This paper is the second in a series that presents a new approach
for numerically implementing the point-splitting regularization scheme
in asymptotically flat spacetimes of black holes, or other compact
objects. Our approach only requires the background to admit some symmetry,
 one which allows (even partial) separation of variables in the field
equation. The first paper \cite{Levi =000026 Ori - 2015 - t splitting regularization}
presented the regularization of $\left\langle \phi^{2}\right\rangle $
in the $t$-splitting variant designed for stationary backgrounds;
here we presented the regularization of this same quantity in the
$\theta$-splitting variant, applicable to spherically symmetric backgrounds.
In both cases we restricted our attention to $\left\langle \phi^{2}\right\rangle $
for simplicity. In forthcoming papers we shall apply our method to
the renormalized stress-energy tensor. We also hope to present a third
variant, azimuthal splitting, which would be usable for generic axially-symmetric
backgrounds. 

To demonstrate how the $\theta$-splitting method works in practice,
we applied it here to the Schwarzschild case, and calculated $\left\langle \phi^{2}\right\rangle _{ren}$
in all three vacuum states: Boulware, Hartle-Hawking, and Unruh. The
results were found to agree very well with our previous results obtained
from $t$-splitting, and also with results calculated by Anderson
using a very different method \cite{Anderson (private)} \textemdash{}
as well as with known analytical results \cite{Candelas - 1979 - phi2 Schwarzschild}
at the horizon and at infinity. 

At the technical level, there is a notable difference between the
$t$-splitting and $\theta$-splitting variants: In the former, as
long as the points are separated (in $t$), the mode-sum operations
are regular. In $\theta$-splitting, by contrast, the integral over
$\omega$ diverges despite the separation in $\theta$. To handle
this intermediate-stage divergence we had to introduced an additional,
auxiliary, separation in $t$ (which we take to vanish before taking
the $\theta$-separation to zero). This makes the $\theta$-splitting
variant slightly more complicated than its $t$-splitting counterpart
\textemdash{} nevertheless not too complicated, as demonstrated by
the Schwarzschild example. 

The analytical processing of the point-separated expression for $\left\langle \phi^{2}\right\rangle _{ren}$
in Sec. \ref{subsec: The-stationary-case} involved a key manipulation:
We have interchanged the order of the $\delta\to0$ limit in Eq. (\ref{eq: Omega integral - phi2 two limits})
with the sum over $l$, and also with the integral over $\omega$.
{[}This interchange led to elimination of the factor $e^{i\omega\delta}$
in the first integral in the R.H.S. of Eq. (\ref{eq: Omega integral - break to two integrals}),
replacing this integral by $F(l)$, hence leading to Eq. (\ref{eq: Omega integral - ph2 final res}).{]}
We are unable to provide a rigorous mathematical proof for the justification
of this interchange. The main obstacle, of course, is the fact that
the functions $\bar{\psi}_{\omega l}(z)$ are not known explicitly.
Nevertheless, we do find strong evidence for the validity of this
interchange. It is fairly clear that the justification (or otherwise)
of such an interchange would primarily depend on the asymptotic behavior
of the function $\left|\bar{\psi}_{\omega l}\left(z\right)\right|^{2}$
at large $l$ and large $\omega$. (For instance, if this function
were vanishing beyond some $l$, the sum would then become a finite
one and its interchange with the $\delta\to0$ limit would be a trivial
operation.) This domain of large $l$ and large $\omega$ is amenable
to WKB analysis. It therefore seems reasonable to assume that for
the sake of addressing this interchangeability issue, one could represent
$\bar{\psi}_{\omega l}\left(z\right)$ by its leading-order WKB approximation.
Preliminary investigation of this issue within leading-order WKB suggests
to us that the $\delta\to0$ limit is indeed interchangeable with
the other operations, justifying our manipulation in Sec. \ref{subsec: The-stationary-case}.
We hope to present the details of this WKB-based analysis elsewhere.
An independent strong evidence for the validity of this operation
comes from the excellent agreement between the results obtained from
$\theta$-splitting and $t$ splitting (and between both of them and
previous results by Anderson \cite{Anderson (private)}). This was
demonstrated in Sec. \ref{subsec:Numerical-implementation-in} for
the Schwarzschild case, and similar agreement was also found in the
Reissner-Nordstrom case. 

We consider the angular-splitting method to be a primary tool in the
investigation of self-consistent semiclassical BH evaporation: Since
an evaporating BH constitutes a time-dependent background, the $t$-splitting
variant is inapplicable to it (at least in the strict direct sense).
Yet the angular-splitting variant should be applicable to this system,
owing to its spherical symmetry. 

\section*{Acknowledgment}

We would like to thank Paul Anderson for sharing his unpublished numerical
results with us, and also for his kind hospitality and fruitful discussions
during our visit at Wake Forest University. This research was supported
by the Asher Fund for Space Research at the Technion. 

\appendix

\section{Generalized sums and integrals\label{sec: Generalized-sums}}

As was already mentioned above, all the infinite sums in this paper
are in principle \textit{generalized sums}, defined according to \textit{Abel
summation method. }Namely, by denoting $\sum_{l=0}^{l=\infty}f\left(l\right)$
we actually mean 

\[
\lim_{\alpha\to0^{+}}\sum_{l=0}^{l=\infty}e^{-\alpha l}f\left(l\right).
\]
The same applies to the integrals over $\omega$: By denoting $\int_{0}^{\infty}f\left(\omega\right)d\omega$
we actually refer to the corresponding \textit{generalized integral}
defined as
\[
\lim_{\alpha\to0^{+}}\int_{0}^{\infty}e^{-\alpha\omega}f\left(\omega\right)d\omega.
\]
The reason for using such generalized sums and integrals is the presence
of undamped oscillations at large $l$ or large $\omega$, as explained
in Sec. \ref{subsec:Calculation-of-:}.

Some of the sums/integral in the paper do converge in the conventional
sense. Note that no ambiguity arises in such cases from the usage
of the same symbol ``$\sum$'' (``$\int$'') for both the conventional
sum (integral) and the generalized one, due to the following consistency
property: Whenever a sum/integral converges in the usual sense, it
is guaranteed to coincide with the corresponding generalized sum/integral.

The oscillations in the summation of various quantities over $l$
usually arise from the oscillatory nature of the Legendre factor $P_{l}\left(cos\varepsilon\right)$.
Therefore these oscillations disappear when $\varepsilon\to0$ is
substituted. Since at the end of the day all the numerical calculations
of the mode functions are carried in this coincidence limit, it follows
that eventually no undamped oscillations in $l$ are encountered in
the numerical evaluation part. (These large-$l$ oscillations only
appear in the preceding, theoretical part of the analysis that involves
the usage of separated $\theta$, namely $\varepsilon\neq0$.) 

The situation with regards to the oscillations at  large $\omega$
is somewhat  different, as we now discuss.

\subsection{Large-$\omega$ oscillations}

The oscillations in the integrals of various quantities over $\omega$
arise from a more geometric reason: The presence of a null geodesic
that connects pairs of points separated in $t$ only. The two-point
function diverges for such null-separated points, leading to large-$\omega$
oscillations in its Fourier decomposition (see \cite{Levi =000026 Ori - 2015 - t splitting regularization}).
Strong ($\propto\omega^{1/2}$) such large-$\omega$ oscillations
were encountered, for example, in $t$ splitting. Unlike the oscillations
in $l$, these $\omega$-oscillations survive even at the limit $t'\to t$.
Therefore, whenever these oscillations occur, they show up even in
the final stage of numerical integration (which is carried after coincidence).
In Ref. \cite{Levi =000026 Ori - 2015 - t splitting regularization}
we described how we practically implement the generalized integral
\textemdash{} and thereby kill the $\omega$-oscillations \textemdash{}
by the ``self-cancellation'' process. 

In the present variant of $\theta$ splitting the large-$\omega$
oscillations are less severe. The main reason is that here we carry
the integral over $\omega$ for each $l$ separately. The quantities
that we integrate are the (square of the absolute value of the) mode
functions, namely solutions of a wave equations, in a fictitious 1+1
spacetime (spanned by the $t,z$ coordinates; and with some $l$-dependent
effective potential). By contrast, in $t$ splitting the $\omega$-integration
is carried \emph{after} summation over $l,m$, hence the relevant
``effective spacetime'' for this manipulation is the true 3+1 spacetime.
Owing to the smaller effective dimension in $\theta$ splitting, the
divergence of the null-connected two-point function is weaker (it
is $\propto\ln\sigma$ compared to $1/\sigma$), and so do the large-$\omega$
oscillations in its Fourier decomposition.

Indeed, in $\theta$-splitting the oscillations in the integrand $\left|\bar{\psi}_{\omega l}\right|^{2}$
decay as $1/\omega$ in the regular-center case {[}see Sec. \ref{subsec:Regular-center},
and particularly the Minkowski example (\ref{eq:Minkowski-2}){]},
and are hence integrable; \footnote{By contrast, in $t$ splitting the oscillations diverge as $\omega^{1/2}$.
\cite{Levi =000026 Ori - 2015 - t splitting regularization}} And in the eternal case there are no large-$\omega$ oscillations
at all. Therefore, self-cancellation of large-$\omega$ oscillations
is not compulsory in $\theta$-splitting. (However, in the regular-center
case, due to the $\propto\omega^{-1}$ oscillations the integral over
$\omega$ converges rather slowly, and self-cancellation of oscillations
is practically needed to speed the convergence.) 

Finally we briefly address the reason, from the geometrical view-point,
for the presence of large-$\omega$ oscillations in the regular-center
case, and their absence in the eternal case. As already mentioned
above, this may be related to the presence (or otherwise) of null
geodesics connecting pairs of points separated by $t$ only, \emph{in
the effective 1+1 spacetime} spanned by $t$ and $z$ (which is the
relevant effective spacetime for the $\omega$-integration in $\theta$
splitting). 

We first need to recall that in 1+1 dimensions, genuine connecting
null geodesics do not exist at all. This follows immediately from
the timelike character of the $t$-separation, combined with the trivial
nature of the light cone in 2d spacetimes. This explains the lack
of oscillations in the eternal case. 

The situation in backgrounds with a regular center is more delicate,
however: Recall that in the effective 1+1 spacetime the high-frequency
wave packets (which usually propagate along null geodesics) actually
bounce when they hit the origin. Hence, the relevant orbits in the
geometrical-optics limit are the ``broken'' null geodesics, which
bounce at the origin. The existence of such ``broken connecting null
geodesics'' (for pairs of points separated in $t$) leads to large-$\omega$
oscillations in the regular-center case. 

Nevertheless, note that there is exactly one such ``broken connecting
null geodesic'' \textemdash{} and hence exactly one oscillation frequency
\textemdash{} at each point $z$. This is illustrated, for example,
in the Minkowski case (\ref{eq:Minkowski-2}). (For comparison, in
$t$ splitting in e.g. Schwarzschild background there is an infinite
discrete set of connecting null geodesics, and hence infinite set
of oscillation frequencies, at each point. \cite{Levi =000026 Ori - 2015 - t splitting regularization})

\section{Legendre blind spots \label{sec: Legendre-sums}}

In this Appendix we prove that the generalized sum 
\[
\sum_{l=0}^{\infty}\left(2l+1\right)\left[l\left(l+1\right)\right]^{n}P_{l}\left(cos\varepsilon\right)
\]
vanishes for any integer $n\ge0$. 

We first treat the $n=0$ case, namely we prove that 
\begin{equation}
\sum_{l=0}^{\infty}\left(2l+1\right)P_{l}\left(cos\varepsilon\right)=0.\label{eq: App Legendre - Lpl zero}
\end{equation}
It is helpful to use the generating function $G\left(\varepsilon,t\right)$
(see Ref. \cite{Mathematical Methods for Physicists}) which for any
finite $\varepsilon$ ($\varepsilon\neq n\pi$) takes the form 
\begin{equation}
\sum_{l=0}^{\infty}P_{l}\left(cos\varepsilon\right)t^{l}=\frac{1}{\sqrt{1-2tcos\varepsilon+t^{2}}}\equiv G\left(\varepsilon,t\right).\label{eq: Legendre generating func.}
\end{equation}
This sum is a \emph{conventional} one (and the same applies to all
sums up to Eq. (\ref{eq:conven}) inclusive); and it converges uniformly
throughout the range $0<t<1$, which we consider here. Differentiating
both sides with respect to $t$ and then multiplying by $t$ we get
\[
\sum_{l=0}^{\infty}l\,P_{l}\left(cos\varepsilon\right)t^{l}=t\,\frac{\partial G\left(\varepsilon,t\right)}{\partial t}=\frac{t\,cos\varepsilon-t^{2}}{\left(1-2tcos\varepsilon+t^{2}\right)^{3/2}}\,.
\]
We can now take a combination of the last two equations: 
\[
\sum_{l=0}^{\infty}\left(2l+1\right)P_{l}\left(cos\varepsilon\right)t^{l}=2t\frac{\partial G\left(\varepsilon,t\right)}{\partial t}+G\left(\varepsilon,t\right)=\frac{1-t^{2}}{\left(1-2tcos\varepsilon+t^{2}\right)^{3/2}}\,.
\]
Next we define $\alpha\equiv-\ln t$, noting that $\alpha>0$ in the
relevant domain. Substituting $t=e^{-\alpha}$ in the last equation
yields 
\[
\sum_{l=0}^{\infty}\left(2l+1\right)P_{l}\left(cos\varepsilon\right)e^{-\alpha l}=\frac{1-e^{-2\alpha}}{\left(1-2cos\varepsilon\,e^{-\alpha}+e^{-2\alpha}\right)^{3/2}}\,.
\]
Taking the limit $\alpha\to0^{+}$ on both sides we find 
\begin{equation}
\lim_{\alpha\to0^{+}}\sum_{l=0}^{\infty}\left(2l+1\right)P_{l}\left(cos\varepsilon\right)e^{-\alpha l}=0\,.\label{eq:conven}
\end{equation}
But this is exactly the definition of the generalized sum in Eq. (\ref{eq: App Legendre - Lpl zero}).
Q.E.D.

Next, we want to prove that
\[
\sum_{l=0}^{\infty}\left(2l+1\right)\left[l\left(l+1\right)\right]^{n}P_{l}\left(cos\varepsilon\right)=0,
\]
for every $n\in\mathbb{N}$. This is easily done by induction, first
assuming that for some $n=k$ 
\[
\sum_{l=0}^{\infty}\left(2l+1\right)\left[l\left(l+1\right)\right]^{k}P_{l}\left(cos\varepsilon\right)=0,
\]
and proving the equality holds for $n=k+1$ as well \textemdash{}
since we already know it is true for $n=0$. The differential equation
that defines the Legendre polynomials can be written (see Ref. \cite{Mathematical Methods for Physicists})
in the form 
\[
l\left(l+1\right)P_{l}\left(\cos\varepsilon\right)=-\sin^{2}\varepsilon\frac{\partial^{2}P_{l}\left(cos\varepsilon\right)}{\left(\partial cos\varepsilon\right)^{2}}+2\cos\varepsilon\frac{\partial P_{l}\left(cos\varepsilon\right)}{\partial cos\varepsilon},
\]
hence
\begin{gather*}
\sum_{l=0}^{\infty}\left(2l+1\right)\left[l\left(l+1\right)\right]^{k+1}P_{l}\left(cos\varepsilon\right)\\
=\sum_{l=0}^{\infty}\left(2l+1\right)\left[l\left(l+1\right)\right]^{k}\left[-\sin^{2}\varepsilon\frac{\partial^{2}P_{l}\left(cos\varepsilon\right)}{\left(\partial cos\varepsilon\right)^{2}}+2\cos\varepsilon\frac{\partial P_{l}\left(cos\varepsilon\right)}{\partial cos\varepsilon}\right]=\\
=\left[-\sin^{2}\varepsilon\frac{\partial^{2}}{\left(\partial cos\varepsilon\right)^{2}}+2\cos\varepsilon\frac{\partial}{\partial cos\varepsilon}\right]\sum_{l=0}^{\infty}\left(2l+1\right)\left[l\left(l+1\right)\right]^{k}P_{l}\left(cos\varepsilon\right)=0.
\end{gather*}
Q.E.D.

\section{Legendre decomposition of the counter-term\label{sec: Legendre-decomposition}}

This Appendix deals with the Legendre decomposition of the counter-term
as given in Eq. (\ref{eq: Omega integral - Gds(epsilon)}). Namely,
we obtain the decompositions (\ref{eq:sin}) and (\ref{eq:LogSin})
for the functions $\sin^{-2}\left(\varepsilon/2\right)$ and $\ln\left[\sin\left(\varepsilon/2\right)\right]$
respectively. 

First we treat the term $\sin^{-2}\left(\varepsilon/2\right)$, and
to this end we define
\begin{equation}
S\left(\alpha\right)\equiv\sum_{l=0}^{\infty}\frac{2l+1}{2}h\left(l\right)P_{l}\left(z\right)e^{-\alpha l}=\sum_{l=1}^{\infty}\frac{2l+1}{2}h\left(l\right)P_{l}\left(z\right)e^{-\alpha l}\,,\label{eq:S_alpha}
\end{equation}
where $h(l)$ is the harmonic number. {[}Here and throughout this
Appendix all sums are conventional ones, except in Eq. (\ref{eq:final}).{]}
Using the Legendre identity \cite{Mathematical Methods for Physicists}
\[
\left(2l+1\right)z\,P_{l}\left(z\right)=\left(l+1\right)P_{l+1}\left(z\right)+lP_{l-1}\left(z\right),
\]
one can write $2zS\left(\alpha\right)$ as
\[
2zS\left(\alpha\right)=\sum_{l=1}^{\infty}h\left(l\right)\left(l+1\right)P_{l+1}\left(z\right)e^{-\alpha l}+\sum_{l=1}^{\infty}h\left(l\right)lP_{l-1}\left(z\right)e^{-\alpha l}\,.
\]
Renaming the $l$ index so as to retain $P_{l}$ (rather than $P_{l\pm1}$)
in both sums, we obtain
\[
2zS\left(\alpha\right)=e^{\alpha}\sum_{l=2}^{\infty}h\left(l-1\right)lP_{l}\left(z\right)e^{-\alpha l}+e^{-\alpha}\sum_{l=0}^{\infty}h\left(l+1\right)\left(l+1\right)P_{l}\left(z\right)e^{-\alpha l}\,.
\]
However, we want to start the summation at $l=1$ (instead of $2$
or $0$) in both sums. In the first sum this change is free because
$h\left(0\right)=0$, but in the second sum we must compensate it
by adding the $l=0$ contribution which amounts to $e^{-\alpha}$:
\[
2zS\left(\alpha\right)=e^{\alpha}\sum_{l=1}^{\infty}h\left(l-1\right)lP_{l}\left(z\right)e^{-\alpha l}+e^{-\alpha}\sum_{l=1}^{\infty}h\left(l+1\right)\left(l+1\right)P_{l}\left(z\right)e^{-\alpha l}+e^{-\alpha}\,.
\]
We now re-express the two sums in terms of $h(l)$ rather than $h\left(l\pm1\right)$:
\[
2zS\left(\alpha\right)=e^{\alpha}\sum_{l=1}^{\infty}\left[h\left(l\right)-\frac{1}{l}\right]lP_{l}\left(z\right)e^{-\alpha l}+e^{-\alpha}\sum_{l=1}^{\infty}\left[h\left(l\right)+\frac{1}{l+1}\right]\left(l+1\right)P_{l}\left(z\right)e^{-\alpha l}+e^{-\alpha}\,,
\]
which we recast as 
\begin{equation}
2zS\left(\alpha\right)=\sum_{l=1}^{\infty}h\left(l\right)\left[le^{\alpha}+\left(l+1\right)e^{-\alpha}\right]P_{l}\left(z\right)e^{-\alpha l}-2\sinh(\alpha)\sum_{l=1}^{\infty}P_{l}\left(z\right)e^{-\alpha l}+e^{-\alpha}\,.\label{eq:TwoSums}
\end{equation}

Let us elaborate on the first sum in Eq. (\ref{eq:TwoSums}). We write
it in the form 
\[
\sum_{l=1}^{\infty}h\left(l\right)\left[\left(2l+1\right)\cosh(\alpha)-\sinh\alpha\right]P_{l}\left(z\right)e^{-\alpha l}\,,
\]
and using the definition of $S\left(\alpha\right)$ we re-express
this sum as 
\begin{equation}
2\cosh(\alpha)S\left(\alpha\right)-\sinh(\alpha)\sum_{l=1}^{\infty}h\left(l\right)P_{l}\left(z\right)e^{-\alpha l}\,\,.\label{eq:FirstSum}
\end{equation}
The second sum in Eq. (\ref{eq:TwoSums}) can be directly computed
(for $|z|<1$ and $\alpha>0$) by setting $\cos\varepsilon=z$ and
$t=e^{-\alpha}$ in the generating function (\ref{eq: Legendre generating func.}):
\[
\sum_{l=1}^{\infty}P_{l}\left(z\right)e^{-\alpha l}=\frac{1}{\sqrt{1-2ze^{-\alpha}+e^{-2\alpha}}}-1\,\,.
\]
Substituting this back in Eq. (\ref{eq:TwoSums}), along with the
expression (\ref{eq:FirstSum}) for the first sum, we obtain 
\begin{equation}
2zS\left(\alpha\right)=2\cosh(\alpha)S\left(\alpha\right)-\sinh(\alpha)\sum_{l=1}^{\infty}h\left(l\right)P_{l}\left(z\right)e^{-\alpha l}+e^{\alpha}-\frac{2\sinh(\alpha)}{\sqrt{1-2ze^{-\alpha}+e^{-2\alpha}}}\,.\label{eq:Long}
\end{equation}
Consider now the limit $\alpha\to0^{+}$ of this equation. The last
term vanishes (recall that we consider here $z<1$). Concerning the
term $\propto\sinh(\alpha)$, we assume here that the sum over $l$
does not diverge as $\alpha\to0^{+}$, hence this term vanishes too.
\footnote{It is actually possible to proceed without using this non-divergence
assumption: Denoting the sum in Eq. (\ref{eq:Long}) by $\tilde{S}\left(\alpha\right)$,
notice that $S=\tilde{S}/2-d\tilde{S}/d\alpha$, which allows one
to treat Eq. (\ref{eq:Long}) as a linear ODE for $\tilde{S}\left(\alpha\right)$.
This ODE is solvable, yielding an explicit expression for $\tilde{S}\left(\alpha\right)$
and hence $S\left(\alpha\right)$. We shall not display this expression
here as it is too long. Nevertheless, when the limit $\alpha\to0^{+}$
is taken, we recover Eq. (\ref{eq:Limit}). } Applying this limit to both sides of the equation we now obtain
\[
2z\lim_{\alpha\to0^{+}}S\left(\alpha\right)=2\lim_{\alpha\to0^{+}}S\left(\alpha\right)+1\,.
\]
Setting $z=\cos\varepsilon$ and extracting the desired limit of $S\left(\alpha\right)$
we find 
\begin{equation}
\lim_{\alpha\to0^{+}}S\left(\alpha\right)=\frac{1}{2\left(z-1\right)}=-\frac{1}{2\left(1-\cos\varepsilon\right)}=-\frac{1}{4\sin^{2}\left(\varepsilon/2\right)}\,\,.\label{eq:Limit}
\end{equation}
Recalling Eq. (\ref{eq:S_alpha}), we finally obtain
\begin{equation}
\sum_{l=0}^{\infty}\frac{2l+1}{2}h\left(l\right)P_{l}\left(\cos\varepsilon\right)=-\frac{1}{4\sin^{2}\left(\varepsilon/2\right)}\label{eq:final}
\end{equation}
(this time with a \emph{generalized} sum), thereby recovering Eq.
(\ref{eq:sin}). Q.E.D.

The second Legendre decomposition needed for the counter-term is that
of $\ln\left[\sin\left(\varepsilon/2\right)\right]$. Expressing it
in the form 
\[
\ln\left[\sin\left(\varepsilon/2\right)\right]=\sum_{l=0}^{\infty}\frac{2l+1}{2}\Lambda\left(l\right)P_{l}\left(\cos\varepsilon\right)
\]
(in accord with Eq. (\ref{eq:LogSin})), we need to calculate the
expansion coefficients $\Lambda\left(l\right)$. We define $z=\cos\varepsilon$,
and noting that
\[
\ln\left[\sin\left(\varepsilon/2\right)\right]=\frac{1}{2}\ln\left(\frac{1-z}{2}\right)\,,
\]
the desired coefficients are given by the Legendre integral 
\begin{gather*}
\Lambda\left(l\right)=\int_{-1}^{1}\frac{1}{2}\ln\left(\frac{1-z}{2}\right)P_{l}\left(z\right)dz\,.
\end{gather*}
For $l=0$ this integral is trivial and the result is $\Lambda(0)=-1$.
For $l>0$ one can use the Legendre equation to rewrite it as
\[
\Lambda\left(l\right)=-\frac{1}{2l\left(l+1\right)}\int_{-1}^{1}\ln\left(\frac{1-z}{2}\right)\frac{d}{dz}\left[\left(1-z^{2}\right)\frac{d}{dz}P_{l}\left(z\right)\right]dz\,.
\]
Integrating by parts yields
\[
\Lambda\left(l\right)=-\frac{1}{2l\left(l+1\right)}\int_{-1}^{1}\left(1+z\right)\frac{d}{dz}P_{l}\left(z\right)dz\,,
\]
as the boundary term vanishes. Integrating by parts once again gives
\[
\Lambda\left(l\right)=-\frac{1}{2l\left(l+1\right)}\left[\left[\left(1+z\right)P_{l}\left(z\right)\right]_{-1}^{1}-\int_{-1}^{1}P_{l}\left(z\right)dz\right].
\]
The integral vanishes for all $l>0$, and from the $z=1$ limit of
the first term we are left with
\[
\Lambda\left(l\right)=-\frac{1}{l\left(l+1\right)}\,\,.
\]

We conclude that
\[
\Lambda\left(l\right)=\begin{cases}
-1 & l=0\\
-\frac{1}{l\left(l+1\right)} & l>0
\end{cases}\,\,.
\]
Q.E.D.

\section{Large-$\omega$ expansion\label{sec: Large w approximation}}

In this Appendix we explore the asymptotic behavior of the field modes
$\psi_{\omega l}$ at large $\omega$. More specifically, we expand
these quantities in powers of $1/\omega$ (at fixed $l$). Understanding
this large-$\omega$ asymptotic behavior is necessary in angular splitting
because the integral of $\left|\bar{\psi}_{\omega l}\right|^{2}$
(which is $\propto\left|\psi_{\omega l}\right|^{2}/\omega$) over
$\omega$ diverges; and in order to regularize it we need to subtract
the appropriate large-$\omega$ piece. 

Luckily, the regularization of $\left\langle \phi^{2}\right\rangle $
only requires the leading-order term in the expansion (namely the
term $\propto\omega^{0}$ in $\psi$). \footnote{For the calculation of the renormalized stress-energy tensor one need
to subtract terms up to order $\omega^{-2}$.} A crucial outcome of the expansion below is that this leading-order
term is independent of $l$. This allows us to regularize the $\omega$-integral
by simply subtracting the $l=0$ contribution (see Sec. \ref{subsec:The-integral-over}). 

But the large-$\omega$ expansion has an additional purpose: Even
after the leading term in $\left|\bar{\psi}_{\omega l}\right|^{2}$
has been subtracted, the integral of the remaining piece decays rather
slowly, typically like $\omega^{-2}$. In principle we need to integrate
up to $\omega=\infty$, but in practice we can only carry the numerical
integration up to some finite value $\omega_{max}$. To account for
the missing contribution from $\omega>\omega_{max}$, we replace the
integrand in this large-$\omega$ domain by its power series in $1/\omega$,
up to a sufficient order. For example, in the numerical implementation
in the Schwarzschild case we use the terms up to order $\omega^{-8}$
(see Sec. \ref{subsec:Numerical-implementation-in}). The reminder
decays very rapidly, hence its contribution at $\omega>\omega_{max}$
is negligible. 

The large-$\omega$ expansion looks different in the eternal and regular-center
cases (also in each of these cases there are differences between static
and time-dependent backgrounds). The eternal case is simpler, because
there are no reflections in the large-$\omega$ limit. On the other
hand, in the presence of a regular center waves are fully reflected
off the origin, even for arbitrarily large $\omega$. Owing to interference
of the propagating and reflected pieces, $\left|\bar{\psi}_{\omega l}\right|^{2}$
turns out to be oscillatory in the regular-center case, in contrast
to its monotonic behavior in the eternal case. This makes the large-$\omega$
expansion more complicated in the case of regular center. (And in
both cases, the time-dependent problem is obviously more complicated
than the static one.) 

In what follows we shall describe the analysis in some detail in the
simplest case of static eternal background. This is also the case
that we need to support our numerical analysis in Schwarzschild. We
shall provide here the expansion coefficients up to order $\omega^{-8}$.
The analysis of the three other cases is more lengthy, and we hope
to present it elsewhere. Nevertheless, in the last subsection we shall
summarize the final results in all four cases (namely eternal and
regular center; static and time-dependent), concerning the leading
order term \textemdash{} the term needed for regularizing the integral
over $\omega$. 

\subsection{Static eternal BH background\label{subsec:Static-eternal-BH}}

In an eternal background, for each $\omega$ and $l$ there are two
sets of basis solutions $\psi_{\omega l}^{in}$ and $\psi_{\omega l}^{up}$,
both satisfy the same radial equation (\ref{eq: Stationary case - field eq-1}).
Let $\psi$ stand for either $\psi_{\omega l}^{in}$ or $\psi_{\omega l}^{up}$.
We express its large-$\omega$ asymptotic behavior as
\begin{equation}
\psi(z)=e^{\pm i\omega z}\sum_{k=0}^{\infty}\frac{a_{k}\left(z\right)}{\omega^{k}}+[...]\,,\label{eq:D1}
\end{equation}
where ``$[...]$'' denotes possible terms that decay faster than
any power of $1/\omega$. Inserting Eq. (\ref{eq:D1}) in the radial
equation we obtain a simple recursion relation 
\[
\pm2ia'_{k+1}=-a''_{k}+V_{l}\,a{}_{k}\,.
\]
When applying this expansion to $\psi_{\omega l}^{in}$ and $\psi_{\omega l}^{up}$,
we denote the corresponding coefficients by $a_{k}^{in}$ and $a_{k}^{up}$.
We now need to use the appropriate boundary conditions at the horizon
and infinity. These are given by Eq. (\ref{eq: Etrrnal static - Basic solutions bounadry conditions})
wherein, in the large-$\omega$ domain, we may set the coefficient
$\rho\to0$ (recalling that the reflection coefficient $\rho$ decays
faster than any power of $1/\omega$). This matching tells us at once
that (i) $\psi_{\omega l}^{in}$ and $\psi_{\omega l}^{up}$ are respectively
associated with the ``-'' and ``+'' signs in Eq. (\ref{eq:D1});
(ii) In both of them the leading-order term is $a_{0}^{in}=a_{0}^{up}=1$,
and (iii) for all $k>0$, $a_{k}^{in}$ vanishes at $z\to\infty$
and $a_{k}^{up}$ at $z\to-\infty$. We thus obtain the integrated
recursion relations for $a_{k}^{in}$ and $a_{k}^{up}$:

\[
a_{k+1}^{in}=-\frac{i}{2}\left(a_{k}^{in}\right)^{\prime}+\frac{i}{2}\int_{\infty}^{z}V_{l}\,a_{k}^{in}\,d\bar{z}\,,
\]
\[
a_{k+1}^{up}=\frac{i}{2}\left(a_{k}^{up}\right)^{\prime}-\frac{i}{2}\int_{-\infty}^{z}V_{l}\,a_{k}^{up}\,d\bar{z}\,.
\]
The calculation of $a_{k}^{in}$ and $a_{k}^{up}$ is now straightforward,
for any $k$. Note that the coefficients $a_{k}^{in,up}$ depend on
the functional form of $V_{l}(z)$, usually in a (multi-) integral
manner.

Next we calculate the large-$\omega$ asymptotic behavior of $\left|\psi_{\omega l}^{in}\right|^{2}$
and $\left|\psi_{\omega l}^{up}\right|^{2}$, which we express as
\[
|\psi(z)|^{2}=\sum_{k=0}^{\infty}\frac{b_{k}\left(z\right)}{\omega^{k}}+[...]\,,
\]
again using the generic symbol $\psi$ for either $\psi_{\omega l}^{in}$
or $\psi_{\omega l}^{up}$. After executing all the $z-$integrals,
the resulting coefficients $b_{k}\left(z\right)$ (namely $b_{k}^{in}$
and $b_{k}^{up}$) turn out to be rather simple, and they demonstrate
several surprising features: (i) $b_{k}$ \textit{vanish} for all
odd $k$; (ii) for all even $k$, $b_{k}\left(z\right)$ depends on
$V_{l}(z)$ in a \emph{direct, local manner} {[}in contrast with the
non-local, integral character of the more elementary coefficients
$a_{k}^{in,up}(z)${]}; (iii) The coefficients $b_{k}^{in}\left(z\right)$
and $b_{k}^{up}\left(z\right)$ are \emph{exactly the same}, for any
$k$. (The derivation of these properties (i-iii) is interesting,
but is way beyond the scope of the present paper.) We may therefore
write the large-$\omega$ expansion in the form 
\begin{equation}
|\psi_{\omega l}^{in}(z)|^{2}=|\psi_{\omega l}^{up}(z)|^{2}=\sum_{k=0}^{\infty}\frac{b_{2k}\left(z\right)}{\omega^{2k}}+[...]\,.\label{eq:Eternal}
\end{equation}
We give here the explicit form of the first few $b_{2k}$ coefficients:
\begin{equation}
b_{0}=1\,,\,\,\,b_{2}=\frac{V_{l}}{2}\,,\,\,\,b_{4}=\frac{1}{8}\left(3V_{l}^{2}-V_{l}''\right)\,,\,\,\,b_{6}=\frac{1}{32}\left(V_{l}^{(4)}-10V_{l}V_{l}''-5V_{l}'^{2}+10V_{l}^{3}\right)\,,\label{eq:coefficients246}
\end{equation}
\begin{equation}
b_{8}=\frac{1}{128}\left[35V_{l}^{4}-70V_{l}^{2}V_{l}''+21V_{l}''^{2}+14V_{l}\left(V_{l}^{(4)}-5V_{l}'^{2}\right)+28V_{l}^{(3)}V_{l}'-V_{l}^{(6)}\right]\,\,,\label{eq:coefficients}
\end{equation}
where a superscript $(n)$ denotes $(\partial/\partial z)^{n}$. 

We used these coefficients in the Schwarzschild and RN cases, for
handling the large-$\omega$ domain in the numerical integration of
$\left|\bar{\psi}_{\omega l}\right|^{2}$. In turn, this numerical
calculation confirmed the validity of the expansion (\ref{eq:Eternal})
along with the coefficients (\ref{eq:coefficients246},\ref{eq:coefficients})
(so far in the Schwarzschild and RN cases). 

Finally, recalling the relation (\ref{eq:auxiliary}) between $\psi_{\omega l}$
and $\bar{\psi}_{\omega l}$, we conclude that at leading order both
$|\bar{\psi}_{\omega l}^{in}(z)|^{2}$ and $|\bar{\psi}_{\omega l}^{up}(z)|^{2}$
are equal to $1/(4\pi r^{2}\omega)$ (and are hence independent of
$l$), with corrections $\propto\omega^{-3}$. 

\subsection{Leading order: summary of results\label{subsec:Leading-order:-summary}}

We summarize here (without proof) the main results concerning the
leading-order behavior of $\left|\bar{\Psi}_{\omega l}\left(z\right)\right|$
in the large-$\omega$ expansion. These results include the eternal
and non-eternal cases, for both static and time-dependent backgrounds. 

Recall that in the static case $\bar{\psi}_{\omega l}$ differs from
$\bar{\Psi}_{\omega l}$ by the factor $e^{-i\omega t}$ only, therefore
$|\bar{\Psi}_{\omega l}|=|\bar{\psi}_{\omega l}|$. 

\subsubsection{Eternal BH \label{subsec:Eternal-BH}}

In this case we obtain 

\begin{equation}
|\bar{\Psi}_{\omega l}^{in}(t,z)|^{2}=|\bar{\Psi}_{\omega l}^{up}(t,z)|^{2}=\frac{1}{4\pi r^{2}\omega}+O(\omega^{-3})\,.\label{eq:Eternal-summary}
\end{equation}
We already derived this result in the previous subsection for the
static case, but this relation holds in the time-dependent case as
well. 

\subsubsection{Regular center\label{subsec:Regular-center}}

In this case there is only one mode function $\bar{\Psi}_{\omega l}\left(t,z\right)$
for each $\omega$ and $l$. Its large-$\omega$ asymptotic behavior
is found to be 
\begin{equation}
\left|\bar{\Psi}_{\omega l}\left(t,z\right)\right|^{2}=\frac{1}{2\pi r^{2}\omega}+(...)\,,\label{eq:non-eternal summary}
\end{equation}
where ``$(...)$'' denotes terms whose integral over $\omega$ converges.
These include two types of terms: (i) terms which decay faster than
$1/\omega$, and (ii) oscillatory terms whose amplitude decays as
$1/\omega$ (or faster), and are hence integrable. 

In the static case we can show that the terms of type (i) (the non-oscillatory
terms) decay as $\omega^{-3}$. In the time-dependent case we haven't
yet obtained the specific decay power of this subdominant term, we
can only show it is faster than $1/\omega$. 

A simple interesting example is the Minkowski background (namely 
$\Gamma=1,\,r=z$): In this case the exact solution for $\psi_{\omega l}(r)$
is 
\[
\psi_{\omega l}(r)=\sqrt{2\pi\omega r}J_{l+1/2}\left(\omega r\right)
\]
where $J$ denotes the Bessel function of the first kind. The asymptotic
behavior at large $\omega$ (for fixed $r>0$) is  
\begin{equation}
|\psi_{\omega l}(r)|^{2}=4\sin^{2}(\omega r-l\pi/2)+O(\omega^{-1})\,,\label{eq:Minkowski}
\end{equation}
which yields
\begin{equation}
|\bar{\Psi}_{\omega l}(r)|^{2}=|\bar{\psi}_{\omega l}(r)|^{2}=\frac{1}{2\pi r^{2}\omega}+\frac{(-1)^{l+1}}{2\pi r^{2}\omega}\cos(2r\omega)+O(\omega^{-2})\,,\label{eq:Minkowski-2}
\end{equation}
in agreement with Eq. (\ref{eq:non-eternal summary}).


\begin{thebibliography}{10}
\bibitem{Hawking - Particle creation by black holes} S. W. Hawking,
Commun. Math. Phys. \textbf{43},199 (1975).

\bibitem{Dewitt - Dynamical theory of groups and fields} B. S. DeWitt,
Dynamical Theory of Groups and Fields (Gordon and Breach, New York,
1965).

\bibitem{Christiansen} S. M. Christensen, Phys. Rev. D \textbf{14},
2490 (1976).

\bibitem{Candelas =000026 Howard - 1984 - phi2 Schwrazschild} P.
Candelas and K. W. Howard, Phys. Rev. D \textbf{29}, 1618 (1984).

\bibitem{Howard - 1984 - Tab Schwarzschild} K. W. Howard, Phys. Rev.
D \textbf{30}, 2532 (1984).

\bibitem{Anderson - 1990 - phi2 static spherically symmetric} P.
R. Anderson, Phys. Rev. D \textbf{41}, 1152 (1990).

\bibitem{Anderson - 1995 - Tab static spherically symmetric} P. R.
Anderson, W. A. Hiscock, D. A. Samuel, Phys. Rev. D \textbf{51}, 4337
(1995).

\bibitem{Ottewill - 2008 Kerr with a mirror} See also a more recent
analysis in Kerr background: G. Duffy and A. C. Ottewill, Phys. Rev.
D \textbf{77}, 024007 (2008). They analyzed the renormalized stress-energy
tensor in a portion of a Kerr BH, in a ``Hartle-Hawking like'' state,
by imposing nonphysical boundary conditions using a mirror. 

\bibitem{Bunch =000026 Devies} T.S. Bunch and P. C. W. Davies, Proc.
R. Soc. Lond. A 357 381-394 (1977).

\bibitem{Levi =000026 Ori - 2015 - t splitting regularization} A.
Levi, A. Ori, Phys. Rev. D \textbf{91}, 104028 (2015).

\bibitem{Christensen =000026 Fulling - 1977} S. M. Christensen and
S. A. Fulling, Phys. Rev. D \textbf{15}, 2088 (1977).

\bibitem{Candelas - 1979 - phi2 Schwarzschild} P. Candelas, Phys.
Rev. D \textbf{21}, 2185 (1980).

\bibitem{Mathematical Methods for Physicists} G. B. Arfken and H.
J. Weber, Mathematical Methods for Physicists, 5th edition, (Academic
Press, San Diego, 2001).

\bibitem{Abramowitz =000026 Stegun} M. Abramowitz and I. A. Stegun,
Handbook of Mathematical Functions, 10th edition, (National Bureau
of Standards, Washington, 1972).

\bibitem{Anderson (private)} P. Anderson, private communication.

\bibitem{Kay and Wald - 1991} B. S. Kay and R. M. Wald, Phys. Reps.
207 49 (1991).

\bibitem{Ottewill - 2000} A. C. Ottewill and E. Winstanley, Phys.
Rev. D \textbf{62}, 084018 (2000).
\end{thebibliography}
\end{document}